\documentclass[aip,amsmath,amssymb,preprint]{revtex4-1}

\usepackage{graphicx}
\usepackage{dcolumn}
\usepackage{bm}
\usepackage[mathlines]{lineno}
\usepackage[colorlinks=true,linkcolor=blue]{hyperref}%
\usepackage[utf8]{inputenc}
\usepackage[T1]{fontenc}
\usepackage{newtxtext,newtxmath}
\usepackage{etoolbox}
\usepackage[mathscr]{euscript}
\usepackage{subcaption}
\usepackage{xcolor}

\makeatletter
\def\@email#1#2{%
\endgroup
\patchcmd{\titleblock@produce}
{\frontmatter@RRAPformat}
{\frontmatter@RRAPformat{\produce@RRAP{*#1\href{mailto:#2}{#2}}}\frontmatter@RRAPformat}
{}{}}
\makeatother

\usepackage{xspace}
\newcommand{\openfoam}{\texttt{Open}\nolinebreak\hspace{-.2em}{\color{blue}\Large$\nabla$}\nolinebreak\hspace{-.2em}\texttt{FOAM}\textsuperscript{\textregistered}\xspace}

\DeclareMathOperator*{\argminA}{arg\,min} 

\begin{document}

\preprint{}

\title{High-Precision Surrogate Modeling for Uncertainty Quantification in Complex Slurry Flows}

\author{M. Elkarii}
\affiliation{MINES ParisTech, PSL Research University, CES, Paris, France}
\affiliation{Mohammed VI Polytechnic University (UM6P), College of Computing, Benguerir, Morocco}

\author{R. Boukharfane}
\affiliation{Mohammed VI Polytechnic University (UM6P), College of Computing, Benguerir, Morocco}
\email{radouan.boukharfane@um6p.ma}

\author{N. El Mo{\c{c}}ayd}
\affiliation{Mohammed VI Polytechnic University (UM6P), College of Agriculture and Environmental Sciences, Benguerir, Morocco}

\date{\today}

\begin{abstract}
Slurry transportation via pipelines is essential for global industries, offering efficiency and environmental benefits. Specifically, the precise calibration of physical parameters for transporting raw phosphate material to fertilizer plants is crucial to minimize energy losses and ensure secure operations. Computational fluid dynamics (CFD) is commonly employed to understand solid concentration, velocity distributions, and flow pressure along the pipeline. However, numerical solutions for slurry flows often entail uncertainties from initial and boundary conditions, emphasizing the need for quantification. This study addresses the challenge by proposing a framework that combines proper orthogonal decomposition and polynomial chaos expansions to quantify uncertainties in two-dimensional phosphate slurry flow simulations. The use of surrogate modeling methods, like polynomial chaos expansion, proves effective in reducing computational costs associated with direct stochastic simulations, especially for complex flows with high spatial variability, as observed in phosphate slurries. Numerical results demonstrate the accuracy of the non-intrusive reduction method in reproducing mean and variance distributions. Moreover, the uncertainty quantification analysis shows that the reduced-order model significantly reduces computational costs compared to the full-order model.
\end{abstract}
\maketitle
\section{Introduction}\label{intro}
The use of numerical modeling to simulate flows of slurry in pipelines has significantly increased, enabling effective design through a thorough understanding of slurry flow characteristics.
Various mixtures have undergone computational fluid dynamics (CFD) simulations.
\citet{dixit2022numerical} investigated sand-water slurry flow in a pipe bend, focusing on assessing head loss.
\citet{sontti2023computational} studied a four-phase slurry used in oil sand ore processing, examining the impact of secondary phases on flow characteristics.
\citet{wan2023numerical} analyzed hydrodynamic characteristics and flow patterns of large particles in deep-sea hydraulic lifting pipes, considering factors such as feed concentration, initial velocity, and particle gradation.
\citet{zhou2023eulerian} explored transport in confined spaces, implementing a bimodal size distribution to enhance accuracy.
\citet{joshi2023slurry} investigated the settling patterns of solid particles in a slurry pipeline, focusing on the influence of Prandtl numbers.
These studies highlight the growing application of numerical modeling in analyzing slurry pipelines, providing valuable insights into complex flow behavior.
Yet, uncertainty remains ubiquitous in all those simulations. 
It is worth to recognize the diverse mechanical perspectives employed in modeling slurry transport. Some studies adopt an approach wherein the slurry is treated as a cohesive fluid continuum, often characterized as non-Newtonian; an illustrative example can be found in \cite{el2021data}. Conversely, alternative investigations conceptualize the flow as solid suspended particles evolving within a carrier Newtonian fluid. Simulations conducted based on both modeling strategies are inherently susceptible to uncertainties. However, it is imperative to emphasize that the present study exclusively concentrates on simulations executed utilizing the latter strategy. All simulations performed in this study utilize the well-known open-source solver, OpenFOAM. This choice is motivated by its demonstrated efficacy in simulating such flows, as evidenced by our prior research endeavors (see \cite{elkarii2022cfd}). Furthermore, the software's widespread use across diverse contexts, including industrial applications, supports its selection. In industrial scenarios where data are limited, practical experiences are unfeasible, and flow configurations are seldom predictable, OpenFOAM proves to be a valuable tool. The primary objective of the present work is to assess uncertainties and conduct sensitivity analyses in such contexts, where decisions must be made prior to the validation of the simulation.
There exist various sources of uncertainty when conducting numerical simulations.
For slurry flow, uncertainties arise from limited knowledge of initial and boundary conditions, material properties, and model parameters.
Assumptions made in physical approximations and numerical parameterization also introduce uncertainties.
Current research focuses on evaluating uncertainty in numerical models \citep{ghanem2017handbook,poette2015iterative}.
UQ can be achieved through intrusive approaches that modify deterministic numerical algorithms \citep{despres2013robust}, or non-intrusive techniques employing black box numerical models \citep{gilli2013uncertainty}.
Since solving two-phase Navier-Stokes equations remains a challenging field \citep{diwan2019pollution}, this study focuses on the non-intrusive method to establish a generic framework for UQ in computational slurry flows.
Stochastic simulations, based on a Monte-Carlo (MC) approach \citep{cheng2019improved,vrugt2007treatment}, are commonly used to quantify uncertainty without modifying the primary model.
Those are based on the central limit theorem \citep{geyer1994convergence}.
They rely on a large number of well-characterized random variable samples in a probabilistic space, resulting in the convergence of distributions according to their probability density functions.
While Monte-Carlo sampling methods guarantee convergence, they require a significant number of simulations \citep{faivre2013analyse}.
Which may be unfeasible for a large range of applications due to the computational cost.
To overcome the limitations of traditional MC simulations, numerous approaches have been introduced to perform UQ.  Their aim is to be at least as accurate as MC methods in estimating uncertainty while requiring less computational time.
Meta-models, such as Kriging and Polynomial Chaos Expansion (PCE), are very useful within this context, by emulating the behavior of the full model with reduced computational effort.
Our past results \citep{elkarii2023global} demonstrate that PCE is particularly efficient for quantifying uncertainties for one dimensional slurry flows simulations. Accordingly this surrogate model is chosen in the current study as well. 

The PCE process initiates by building surrogate models known as Polynomial Chaos expansions (PCEs). 
PCE theory utilizes orthogonal polynomials to establish the basis \citep{xiu2002wiener}, enabling the computation of polynomial coefficients that define a reduced model.
The intrusive method modifies the existing code by approximating the model response in the relevant partial differential equations \citep{askey1985some}.
Alternatively, non-intrusive methods treat the code as a black box, obtaining response solutions through model evaluations \citep{askey1985some}.
Both intrusive and non-intrusive approaches derive polynomial approximations using ensemble-based strategies \citep{ghanem1991stochastic}.
Originally proposed as an alternative to Monte Carlo methods, PCE has proven effective in uncertainty quantification and sensitivity analysis \citep{wang2019optimal}.
Generalized Polynomial Chaos (gPC), an extension of PCE, handles probability distributions beyond Gaussian rules using the Askey scheme \citep{xiu2002wiener}.
For convenience, we use the term PCE throughout this study.

Addressing the difficulties posed by the complex numerical models as part of the development of meta-models to lower the computing cost in the UQ has received a lot of interest.
This is mostly because the construction time needed for PCE rises exponentially with the stochastic dimension.
One of the solutions to tackle the problem of stochastic dimensions was to introduce the adaptive sparse PCE.
The objective is to estimate the PCE using a regression approach using least angle regression, as specified in \citep{efron2004least}.
Another way to counter this curse of dimensionality was to combine Kriging with PCE to offer a novel meta-model \citep{yamazaki2018stochastic}.

This method enabled \citep{elkarii2023global} to conduct a UQ and a global sensitivity analysis (GSA) for one-dimensional quantities of interests (QoI)s related to the transportation of phosphate slurry through pipelines.
The objective of this study is to expand the dimensionality of our QoIs from one dimension to two dimensions.
Expanding our study to consider two-dimensional QoIs in UQ and GSA holds significant importance. 
While one-dimensional analyses provide valuable insights into specific local aspects of slurry flow in pipelines, incorporating the second dimension allows for a more comprehensive understanding of the system's behavior.
By considering both spatial dimensions, we can capture additional information about variables such as particle distribution, velocity profiles, which play crucial roles in optimizing slurry transportation processes. 
The inclusion of two-dimensional QoIs enhances our ability to identify and address potential challenges, such as anisotropic behavior, localized flow variations, or concentration gradients that may impact the system's performance.
Additionally, a higher-dimensional analysis facilitates a more accurate representation of the flow phenomena, allowing us to refine the surrogate models. 

The adaptive sparse PCE previously discussed was used in \citep{elkarii2023global}, to alleviate the dimension in the model input parameters, however, they neglect the spatial dimension of the output variables.
Classically when the output solution is a physical field discretized in a mesh, the approach requires PCE calculation at each node of the mesh, as shown in \citep{el2018polynomial}.
In such circumstances, the numerical resolution of complicated slurry flow characteristics necessitates the use of fine meshes, which incurs large computing costs.
In addition, it should be emphasized that the physical outputs are stochastic processes with spatial and temporal descriptions, not only stochastic random variables.
Therefore, outputs are computed for each node in the computational mesh and at every time step in the time interval.
Which necessitates the development of less time-intensive methods for quantifying uncertainty when the output is a stochastic process.
Various approaches, such as the Principal Component Analysis (PCA) \citep{blatman2011principal}, the Proper Generalized Decomposition (PGD) \citep{chevreuil2012model}, and the Proper Orthogonal Decomposition POD \citep{raisee2015non}, might be used to address this problem.
Indeed, the majority of these techniques emanate from the domain of data science.
Specifically, they fall under the category of unsupervised learning techniques, facilitating clustering analysis.
In the context of uncertainty quantification, these clusters represent distinct physical patterns driven by stochastic modes—those that are the drivers of uncertainty.
While their utilization has witnessed a growing trend within the field, the systematic assessment of their application to intricate physical problems, such as slurry flows, remains an unexplored territory until now.
It is also important to recognize that these strategies operate under the assumption that uncertainties can be captured by a few non-physical factors manifested in the response of random vectors.
By resampling the newly constructed surrogate model, the relevant statistical quantities can be obtained.

Due to the complex physical characteristics of slurry flow, including steep gradients and localized structures in response to the two-phase Navier-Stokes equations, this study employs the POD technique.
Our research introduces a non-intrusive reduced-order approach for simulating the two-dimensional transport of slurries.
The main focus of this study is to analyze the uncertainty associated with initial and boundary conditions, as well as solid properties, and its impact on the energy efficiency of slurry pipe flows.
Random environmental factors often affect the slurry transport energy consumption, leading to uncertainties in measured results. Quantifying these uncertainties is important to validate deterministic solutions.
The specific energy consumption ($\textit{SEC}$), which measures the energy required to move a slurry batch over a given distance, is used to quantify the transport efficiency.
It is directly proportional to the hydraulic gradient, representing the drop in static pressure due to friction in the slurry flow per unit length of the pipe.
The proposed methodology focuses on the UQ in terms of pressure drop and delivered solid concentration, resulting from uncertain modeling and physical parameters.
Additionally, a GSA is conducted using a variance-based method, specifically Sobol' indices.
To assess the performance of the proposed methodology, a comparison is made with a class of extensive Monte Carlo simulations in terms of mean and variance fields.

The remaining sections of this paper are organized as follows.
Section \ref{CFD} presents the slurry transport model, the CFD modeling procedure, and the numerical validation conducted.
In Section \ref{PCE}, the model reduction methodology is discussed, along with the PCE and POD techniques employed.
Section \ref{POD-results} presents the results obtained for UQ and sensitivity analysis of the QoIs.
Finally, Section \ref{general conclusion} provides concluding remarks based on the findings of this study.
%

\section{Uncertainty quantification in stochastic processes}
\label{PCE}
The overall concept is described with a view toward addressing stochastic slurry pipe flows.
Polynomial chaos expansions and stochastic proper orthogonal decomposition techniques are briefly discussed.
The suggested UQ approaches are also employed to calculate sensitivity indices in the present study.
%
\subsection{Polynomial Chaos Expansion} 
%
Within the framework of uncertainty quantification, the PCE has seen extensive application as a surrogate model, 
as its objective is to recreate the global behavior of a simulation following a polynomial decomposition.
The latter are multivariate orthogonal polynomials and serve as the basis functions selected with respect to the joint probability distributions of the stochastic input variables following the so-called Askey scheme of polynomial \citep{xiu2002wiener}.
In the present study, if our variables are stored in the vector \(\mathbf{X} \in \mathbb{R}^{d}\) where $d$ is the number of input variables in $\mathbf{X}$, and \(\mathcal{Y}\) represents the model responses discussed earlier, then one may write, according to PCE:
\begin{equation}
\mathcal{Y}=\mathcal{M}(\mathbf{X})=\sum_{\mathbf{\alpha}\in \mathbb{N}^{d}} \mathbf{a}_{\mathbf{\alpha}} \psi_{\mathbf{\alpha}}(\mathbf{X}),
\label{PC expansion}
\end{equation}
where $\mathbf{\alpha}=\left\{\alpha_1,\ldots\alpha_d\right\}$ denotes a $d$-dimensional index, $\left\{\mathbf{a}_{\alpha},\mathbf{\alpha}\in\mathbb{N}^d\right\}$ are the PCE coefficients to be computed.
The $\psi_{\mathbf{\alpha}}$ are the multivariate polynomial basis elements of degree $|\mathbf{\alpha}|=\sum_{i=1}^{d}\alpha_i$ that are orthonormal with regards to the joint pdf, \(f_{\mathbf{X}}\), of \(\mathbf{X}\), \textit{i.e.}, \(\mathbb{E}\left[\psi_{\mathbf{\alpha}}(\mathbf{X}) \psi_{\mathbf{\beta}}(\mathbf{X})\right]=1\) if \(\mathbf{\alpha}=\mathbf{\beta}\) and 0 otherwise.
For practical purposes, the determination of a PCE is dependent upon the calculation of the spectral coefficients $\left\{\mathbf{a}_{\alpha}, \alpha \in \mathcal{A}\right\}$ where $\mathcal{A}\subset\mathbb{N}^d$ is a finite set of multi-indices such that
\begin{equation}
\mathcal{Y}\approx \mathcal{M}^{\text{PC}}\left(\mathbf{X}\right)=\sum_{\boldsymbol{\alpha}\in\mathcal{A}}\mathbf{a}_{\boldsymbol{\alpha}}\psi_{\boldsymbol{\alpha}}\left(\mathbf{X}\right).
 \end{equation} 
To construct such a set, a standard truncation scheme, which corresponds to all polynomials in the $d$ input variables of total degree less than or equal to $p$ is defined by the set $\mathcal{A}^{d,p}$.
This set is traditionally defined as $\mathcal{A}^{d, p}=\left\{\alpha \in \mathbb{N}^{d}:|\alpha| \leq p\right\}$.
It is worth noting that when this method is used to define the set of multi-indices, \(\mathcal{A}^{d, p}\) comprises $(d+p)!/d! p!$ items following the binomial rule for a given value of the maximum polynomial degree $p$.
If the model's response is strongly nonlinear with numerous input parameters (requiring a large value of $p$) or if the input vector $X$ is large (\text{e.g.}, $M > 10$), implementing such a truncation scheme would result in challenging problems.
This issue is commonly referred to as the curse of dimensionality, we refer the reader to \citep{xiu2003modeling,xiu2010numerical} for further details. 
In the current study, for the purpose of simplicity and without losing generality, just the regression approach is studied, and the most efficient sparse decomposition may be used when the stochastic dimension is very large \citep{blatman2011adaptive}.
It must be emphasized that the well-established quadrature methods are not examined in the present work since they cannot be utilized to construct POD modes.
To estimate the coefficients $\mathbf{a}=\left\{\mathbf{a}_{\mathbf{\alpha}}\in\mathcal{A}\right\}$, the regression approach relies on the solution of a least-square (LS) minimization problem in any $\ell_2$-norm \citep{choi2004polynomial,berveiller2006stochastic}.
To put this into application, the expansion coefficients $\mathbf{a}$ are calculated as:
\begin{equation}
\mathbf{a}=\argminA_{\mathbf{a}\in\mathbb{R}^{\mathtt{M}}}\mathbb{E}\left[\left(\mathcal{Y}-\sum_{\mathbf{\alpha}\in \mathcal{A}} \mathbf{a}_{\mathbf{\alpha}} \psi_{\mathbf{\alpha}}(\mathbf{X})\right)^2\right],
\end{equation}
where $\mathtt{M}$ the cardinal of the set $\mathcal{A}$. The error is defined as the difference between the (exact) model's evaluation 
\begin{equation}
\mathcal{Y}=\left\{\mathcal{M}\left(X_1\right),\ldots,\mathcal{M}\left(X_{\mathtt{N}}\right)\right\},
\end{equation}
and the PCE surrogate estimations for a finite training set or experimental design (DE) of $\mathtt{N}$ randomly sampled input varfiables $\mathbf{X}=\left\{X_1,\ldots,X_{\mathtt{N}}\right\}$ where $X_i\in\mathbb{R}^d$.
The set of $\mathtt{N}$ realization of the input vector, $\mathbf{X}$, is then needed, called design of experiment (DE). 
The generation of the DE is obtained in the present work using the Latin Hypercube sampling approach, which helps to reduce the computational cost of the classical Monte Carlo sampling \citep{iman2008atin}. 
The LS error is transformed into a discretized mean-square problem as:
\begin{equation}
\widehat{\mathbf{a}}=\argminA_{\mathbf{a}\in\mathcal{R}^{\mathtt{M}}}\frac{1}{\mathtt{N}}\sum_{i=1}^{\mathtt{N}}\left(\mathcal{Y}^{(i)}-\sum_{\mathbf{\alpha}\in \mathcal{A}} \mathbf{a}_{\mathbf{\alpha}} \psi_{\mathbf{\alpha}}\left(X_i\right)\right)^2.
\label{LS error}
\end{equation}
Following the work of \citet{blatman2009adaptive}, the regression estimates of the polynomial chaos coefficients are given by:
\begin{equation}
\widehat{\mathbf{a}}=\left(\boldsymbol{\Psi}^\intercal\boldsymbol{\Psi}\right)^{-1}\boldsymbol{\Psi}^\intercal\mathcal{Y},
\end{equation}
where $\boldsymbol{\Psi}$ is the information matrix of size $\mathtt{N}\times \mathtt{M}$ give by $\Psi_{ij}=\psi_j\left(X_i\right)$ assembling the values of all orthonormal polynomials at the ED nodes. 
In case of high dimensional problem an attractive modification on the conventional least squares formulation could be adopted, that limits the sum of the absolute regression coefficients.
In fact, the least-angle regression (LAR), offers a sparse PCE representation with fewer regressors than the traditional complete representation.
Here, we apply the LAR approach as suggested by \citet{blatman2011adaptive}.
For further information on the LAR approach and its use in the context of adaptive sparse PCE, see \citep{efron2004least} and \citep{blatman2011adaptive}, respectively.
Finally, in order to evaluate the quality of the PCE, Leave-One-Out (LOO) error is often used as an excellent measure of accuracy since it enables error estimates without acquiring further forward evaluation of the numerical model \citep{blatman2010adaptive}. 
The definition of the relative LOO error is as follows:
\begin{equation}
\epsilon_{\text{LOO}}=\frac{\sum_{i=1}^{\mathtt{N}}\left(\frac{\mathcal{M}\left(X_i\right)-\mathcal{M}^{PC}\left(X_i\right)}{1-h_{i}}\right)^{2}}{\sum_{i=1}^{\mathtt{N}}\left(\mathcal{M}\left(X_i\right)-\hat{\mu}_{\mathcal{Y}}\right)^{2}},
\label{LOO}
\end{equation}

where \(h_{i}\) is the \(i^{t h}\) diagonal term of \(\mathbf{\Psi}\left(\mathbf{\Psi}^{\intercal} \mathbf{\Psi}\right)^{-1} \mathbf{\Psi}^{\intercal}\) and \(\hat{\mu}_{Y}=\frac{1}{\mathtt{N}} \sum_{i=1}^{\mathtt{N}} \mathcal{M}\left(X_i\right)\)
According to a research summarized by \citet{molinaro2005prediction}, the leave-one-out method has a low estimate bias and a low mean-square error and could be comparable with standard relative $\ell_2$ error in some cases, see for example \citep{el2018polynomial}.
%
\subsection{Stochastic proper orthogonal decomposition}
\label{POD}
%
The POD method, enables a high-dimensional system to be addressed by a low-dimensional one.
This approach involves establishing a set of orthogonal eigenvalues that are representative of the simulated physics.
Solving the integral of Fredholm yields the eigenvectors, while the kernel of this integral is derived from a series of simulations performed using an experimental design.
Specifically, the eigenfunctions associated with the problem are optimal within the context of the dynamic representation (explained below), allowing them to be used to build a simplified representation of physics.
In uncertainty quantification, the POD serves to decrease the size of a random vector produced by the model.
As a result, the uncertainty is applied for each direction given by the eigenvectors $\lambda_{i}$.
The concept is based on the projection of the model's solution $\mathcal{M}\left(X_i\right)$ onto a finite and orthonormal basis \(\left\{\phi_{i}, i \in \mathcal{I}_{\text{POD}}\right\}\), where \(\mathcal{I}_{\text{POD}}\) is a discrete finite collection of indices.
Thus, the procedure $\mathcal{M}\left(X_i\right)$ is decomposed as follows:
\begin{equation}
\mathcal{M}\left({X}\right)=\sum_{i \in \mathcal{I}_{\text{POD}}} \lambda_{i}({X}) \phi_{i},
\label{POD_decomposition}
\end{equation}
where $\phi_{i}$ is estimated by decomposing the spatial covariance matrix constructed as:
\begin{equation}
\mathbf{C}=\frac{1}{\mathtt{N}} \mathcal{M} \mathcal{M}^{\intercal}.
\label{covariance}
\end{equation}
%
%
In addition, we construct a POD-truncated error \(\varepsilon\) so that only the \(k\) most valuable eigenvectors are maintained as the solution following the estimate:
\begin{equation}
\frac{\sum_{i=0}^{k} \lambda_{i}}{\sum_{i=0}^{\mathtt{N}} \lambda_{i}}>1-\varepsilon,
\label{eigenvalues}
\end{equation}
where \(\widehat{\lambda}_{i}\) is the mean value of \(\lambda_{i}({X}_{i})\). 
More details about the POD technique are available in \cite{el2020stochastic}.
%
		
%
%
%
\subsection{POD-PCE meta-model} 
%
The eigenvalues are handled as random variables after rebuilding the stochastic POD.
This suggests that we may preserve the spatial dependence suggested by the eigenvectors $\phi_{i}$ while generating a PCE for each eigenvalue in the manner described in the section on polynomial chaos expansions before.
\begin{equation}
\lambda_{i}({X})=\sum_{j=0}^{\mathtt{M}} \gamma_{j} \psi_{j}\left({X}_{i}\right).
\end{equation}
The equation \eqref{PC expansion} becomes,
\begin{equation}
\mathcal{M}\left({X}\right)=\sum_{i \in \mathcal{I}_{POD}} \left(\sum_{j=0}^{\mathtt{M}} \gamma_{j} \psi_{\alpha}\left({X}\right)\right)\phi_{i}.
\end{equation}
A summary of both algorithms retained in the present study for quantification of uncertainties in stochastic slurry flows is depicted in Fig.~\ref{fig:UQ-flowchart}.
\begin{figure}[ht!]
\centering
\includegraphics[width=\textwidth]{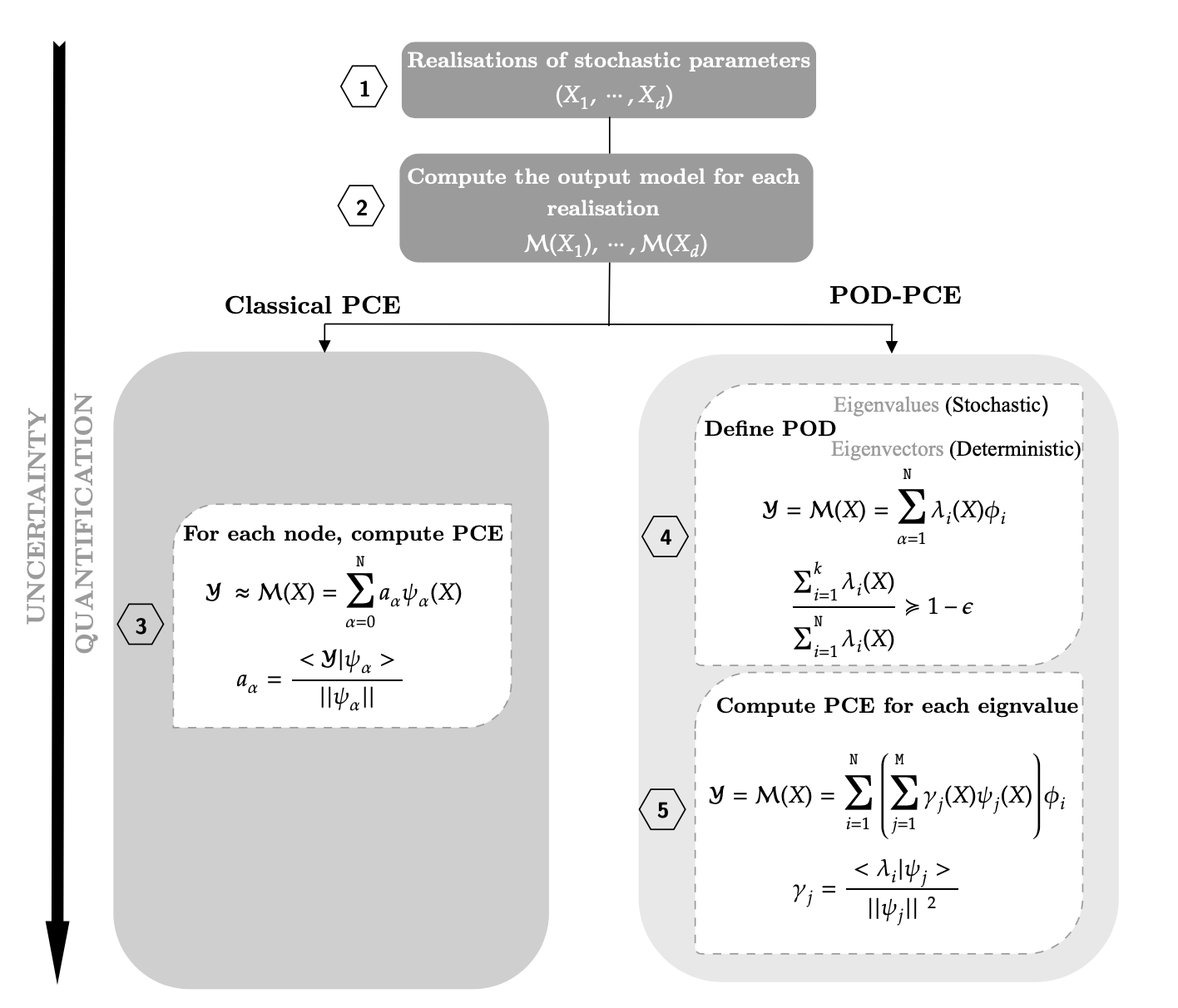}
\caption{Diagram illustrating the distinction between the traditional PCE-based surrogate model and the POD-based surrogate model.}
\label{fig:UQ-flowchart}
\end{figure} 
It should be noted that the traditional method of dealing with an uncertainty quantification issue utilizing the PCE is conducted when a decomposition is completed for each node in the computational mesh.
This procedure is shown in Fig.~\ref{fig:UQ-flowchart} flowchart by steps 1 through 3.
However, as previously stated, one way to ease the spatial distribution is to do a first reduction using the POD.
The stochasticity is assumed to be incorporated in the related eigenvalues, this latter will aid in distinguishing the spatial dependence from the stochastic dependency.
Once the POD is performed, just a few eigenmodes are preserved, allowing for less PCE to be generated than its traditional equivalent.
This procedure is shown in Fig.~\ref{fig:UQ-flowchart} flowchart as stages 1, 2, 4, and 5.
The accuracy of the approaches under consideration is evaluated by comparing their findings to large MC simulations (with N=1000 samples) for the quantities of interest in the mean and variance fields.
This is generally achieved using a $l_2$-error following this: 

\begin{equation}
l_2(\theta)=(\theta_{\text{MC}}-\theta_{\text{POD-PCE}})^2
\end{equation}
where the $\theta$ is the required quantity of interest (Mean value or variance value), $\theta_{\text{MC}}$ refer to its estimation using Monte-Carlo simulations and $\theta_{\text{POD-PCE}}$ refer to its estimation using the surrogate model.
%
\subsection{Aggregated Sobol' indices}
%
The sensitivity indices in this paper are derived using the PCE approach with a non-intrusive sampling procedure.
Therefore, once the PCE is built, the mean $\mu$ and the total variance $D$ can be obtained using the value of the spectral coefficients such that $\mu=a_{0}$ and $D=\sum_{\alpha \in \mathcal{A} \setminus 0} a_{\mathbf{\alpha}}^{2}$.
The computation of Sobol' indices of the output variable $\mathcal{M}_{j}\left({X}_{i}\right)$ concerning the input variable ${X}_{i}$ of any order is a straightforward process.
The first order Sobol' indices can be expressed as follows $\mathtt{S}_{i}^{j}=\sum_{\mathbf{\alpha} \in \mathcal{A}_{i}} a_{\mathbf{\alpha}}^{2}/D$ with $\mathcal{A}_{i}=\left\{\mathbf{\alpha} \in \mathcal{A}: \alpha_{i}>0, \alpha_{j \neq i}=0\right\}$, while the total Sobol' indices is computed as $\mathtt{S}^{T}_{i,j}=\sum_{\mathbf{\alpha} \in \mathcal{A}_{i}^{T}} a_{\mathbf{\alpha}}^{2}/D$ with $\mathcal{A}_{i}^{T}=\left\{\mathbf{\alpha} \in \mathcal{A}: \alpha_{i}>0\right\}$.
Since the QoIs under consideration in the present studies have spatial dependencies,  sensitivity indices should be defined accordingly.
The aggregated first Sobol’ index $GS_{i}$ as introduced in \citep{blatman2009adaptive} are defined here:
\begin{equation}
\mathtt{GS}_{i}=\frac{\sum_{j=1}^{p} \operatorname{Var}\left(\mathcal{M}\left({X}_{j}\right)\right) S_{i}^{j}}{\sum_{j=1}^{p} \operatorname{Var}\left(\mathcal{M}\left({X}_{j}\right)\right)}
\end{equation}
The aggregated total Sobol’ index are calculated as:
\begin{equation}
\mathtt{GS}^{T}_{i}=\frac{\sum_{j=1}^{p} \operatorname{Var}\left(\mathcal{M}\left({X}^{(j)}\right)\right) S^{T}_{i,j}}{\sum_{j=1}^{p} \operatorname{Var}\left(\mathcal{M}\left({X}^{(j)}\right)\right)}
\end{equation}
Notice that \(\mathtt{GS}_{i} \in[0,1]\). As in the case of the scalar indices, the main advantage of the aggregated indices is their easy interpretation: a high value
means the input is important, a value close to zero means it is not, and \(\sum_{i=1}^{N} \mathtt{GS}_{i}=1\).
%
%
\section{Computational Model (CFD modeling procedure)}
\label{CFD}
%

The system under consideration is a three-dimensional pipeline with a diameter of $D=0.103$ m and a length of $L=140D$ to ensure fully developed flow. The transported slurry consists of a mixture of phosphate and water, with densities of $\rho_{s}=2600$ kg/m$^3$ and $\rho_l=1000$ kg/m$^3$, respectively. The actual flow of phosphate slurry operates under specific conditions.
Regarding the flow rate, the pipeline has a capacity of approximately $4400$ t/h, depending on the properties of the slurry. This corresponds to a mixture inlet velocity of $u_{m} = 2.16$ m/s, and it can accommodate a maximum deposit limit velocity of ${u}_{dl} = 1.75$ m/s. The delivered phosphate pulp has a volumetric solid concentration of approximately $\phi_{s} = 33.3\%$, and it contains rigid solid particles with random shapes, having equivalent spherical diameters ranging from $44~\mathrm{\mu}$m to $250~\mathrm{\mu}$m.
For further information about the transportation process and the operating limits of the pipeline, refer to the work of \citet{rusconi2016design}.
%

\subsection{CFD numerical modeling}
%

The two-phase flow is modeled using the Eulerian-Eulerian approach implemented in the \texttt{twoPhaseEulerFoam} solver available in \texttt{OpenFOAM}~\citep{weller1998tensorial}.
The mass and momentum balance equations for the liquid ($k\equiv l$) and solid ($k\equiv s$) phases are solved with closure models for the turbulent stress tensor and interphase forces.
The mass and momentum balance equations are the following:
\begin{equation}
\frac{\partial \rho_k \phi_k}{\partial t} + \nabla \cdot \left( \rho_k \phi_k \mathbf{u}_k \right) = 0,
\end{equation}
\begin{equation}
\frac{\partial \rho_k \phi_k \mathbf{u}_k}{\partial t} + \nabla \cdot \left( \rho_k \phi_k \mathbf{u}_k \mathbf{u}_k \right) = - \nabla \cdot \left( \phi_k \mathbf{\tau}_k \right) - \phi_k \nabla p + \phi_k\rho_k\mathbf{g} + \mathbf{F}_k,
\end{equation}
where $\rho$ is the density, $\alpha$ is the phase fraction, $\mathbf{u}$ is the average velocity, $p$ is the average pressure field shared by all the phases, and $\mathbf{g}$ is the gravity field.
The inter-phase momentum transfer term is denoted as $\mathbf{F}_k$, accounting for the drag, lift, virtual mass, and turbulent drag forces.
For the dispersed solid phase, the inter-phase momentum transfer term takes the following form:
\begin{equation}
\mathbf{F}_s = \mathbf{F}_{d,s} + C_l \rho_l \mathbf{u}_r \times \left( \nabla \times \mathbf{u}_c \right) + C_{vm} \rho_l \left( \frac{D_l \mathbf{u}_l}{Dt} - \frac{D_s \mathbf{u}_s}{Dt} \right) - C_{td} \alpha_s \rho_c \kappa_m \nabla \alpha_s,\label{eq::interphaseForces}
\end{equation}
Here, $\kappa_m$ represents the multiphase mixture of turbulent kinetic energy. The interfacial momentum balance implies that $\mathbf{F}_s = -\mathbf{F}_l$.
In Eq.~\eqref{eq::interphaseForces}, $\mathbf{F}_{d,s}$ is the drag force acting in the solid phase.
The symmetric drag formulation proposed by Weller~\citep{weller2002derivation} is applied:
\begin{equation}
\mathbf{F}_{d,s} = \alpha_s \alpha_l \left( f_s \frac{C_{d,l} \rho_s}{d_s} + f_l \frac{C_{d,s} \rho_l}{d_l} \right) |\mathbf{u}_r| \mathbf{u}_r,
\end{equation}
where
\begin{equation*}
f_k = \min \left[ \max \left( \frac{\phi_k - \phi_k^{\max,f}}{\phi_k^{\max,p} - \phi_k^{\max,f}}, 0 \right), 1 \right],
\end{equation*}
where $\alpha_l^{\max,p} = \alpha_s^{\max,p} = 0.63$. In this formulation, $\mathbf{u}_r = \mathbf{u}_s - \mathbf{u}_l$ represents the relative velocity between phases. The coefficients $C_d$, $C_l$, $C{vm}$, and $C{td}$ correspond to the drag, lift, virtual mass, and turbulent drag coefficients, respectively, with values set as $C_l=C_{vm}=C_{td}=0.5$.
The drag coefficients are determined using the Schiller and Naumann correlation~\citep{schiller1933uber,schiller1933drag}:
\begin{equation*}
C_{d,k} = \max\left[ 0.44, \frac{24}{\mathrm{Re}_k} \left( 1 + 0.15 \mathrm{Re}_{k}^{0.687} \right) \right],~
\end{equation*}
where the Reynolds numbers based on the solid and liquid phase diameters, $\mathrm{Re}_s$ and $\mathrm{Re}_l$ respectively, are defined as $\mathrm{Re_{s}} = \rho_{l} \mathbf{u}_{r} d_s/\mu_l$ and $\mathrm{Re_{l}} = \rho_{l} \mathbf{u}_{r} d_l/\mu_l$, where $d$ represents the dispersed phase diameter and $\mu$ is the dynamic viscosity.
The total stress tensor per unit mass, denoted as $\mathbf{\tau}_k$, is calculated using a Reynolds-Averaged Navier-Stokes approach for the turbulence modeling of the two-phase mixture, as described in \citep{behzadi2004modelling}:
\begin{equation}
\mathbf{\tau}_k = \mu_{\mathrm{eff},k}\left[ \nabla\mathbf{u}_k + \left(\nabla \mathbf{u}_k\right)^\intercal - \frac{2}{3} \mathbf{I} \nabla\cdot\mathbf{u}_k \right],
\end{equation}
where $\mu_{\mathrm{eff},k}$ represents the effective phase viscosity, combining the molecular and turbulent dynamic viscosity of the phase, given by $\mu_{\mathrm{eff},k} = \mu_{k} + \mu_{t,k}$.
However, in relation to the viscosity of the solid phase, it's necessary to use closures to represent how the behavior of this phase, encompassing aspects like contact pressure and velocity collisions, translates into continuous pressure and viscosity.
This study employs the Kinetic Theory of Granular Flow (KTGF), which establishes a relationship between the overall kinetic energy of a group of particles moving randomly within the carrier fluid and their varying velocity, employing the concept of granular temperature.
The viscosity coefficients encompass a combination of three distinct elements:
\begin{equation}
\mu_s=\mu_{s,\mathrm{col}}+\mu_{s,\mathrm{kin}}+\mu_{s,\mathrm{fric}},
\end{equation}
where \(\mu_{s,\mathrm{col}}\), \(\mu_{s,\mathrm{kin}}\), and \(\mu_{s,\mathrm{fric}}\) represent the collisional \cite{gidaspow1991hydrodynamics}, kinetic \cite{syamlal1993mfix}, and frictional \cite{schaeffer1987instability} components of the total shear stress, respectively, as derived from KTGF theory. 
Further elaboration on the formulation of each term can be found in \cite{ding1990bubbling,elkarii2022cfd}.

\begin{figure}[ht!]
\centering
\subfloat[\label{fig:3D domain}]{\includegraphics[width=0.45\textwidth]{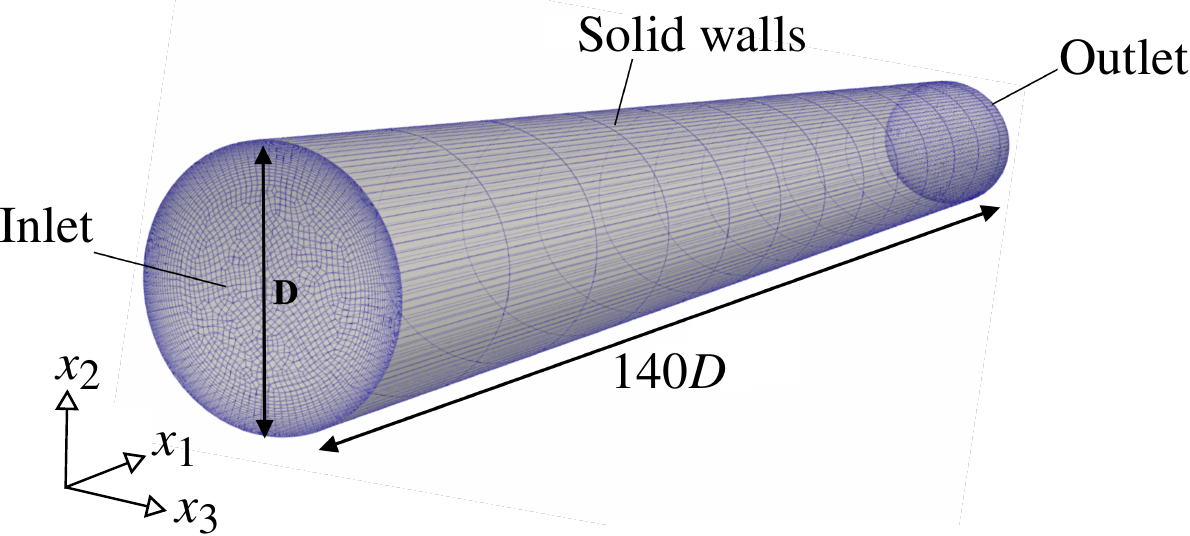}}\hfill
\subfloat[\label{fig:Section_mesh}]{\includegraphics[width=0.45\textwidth]{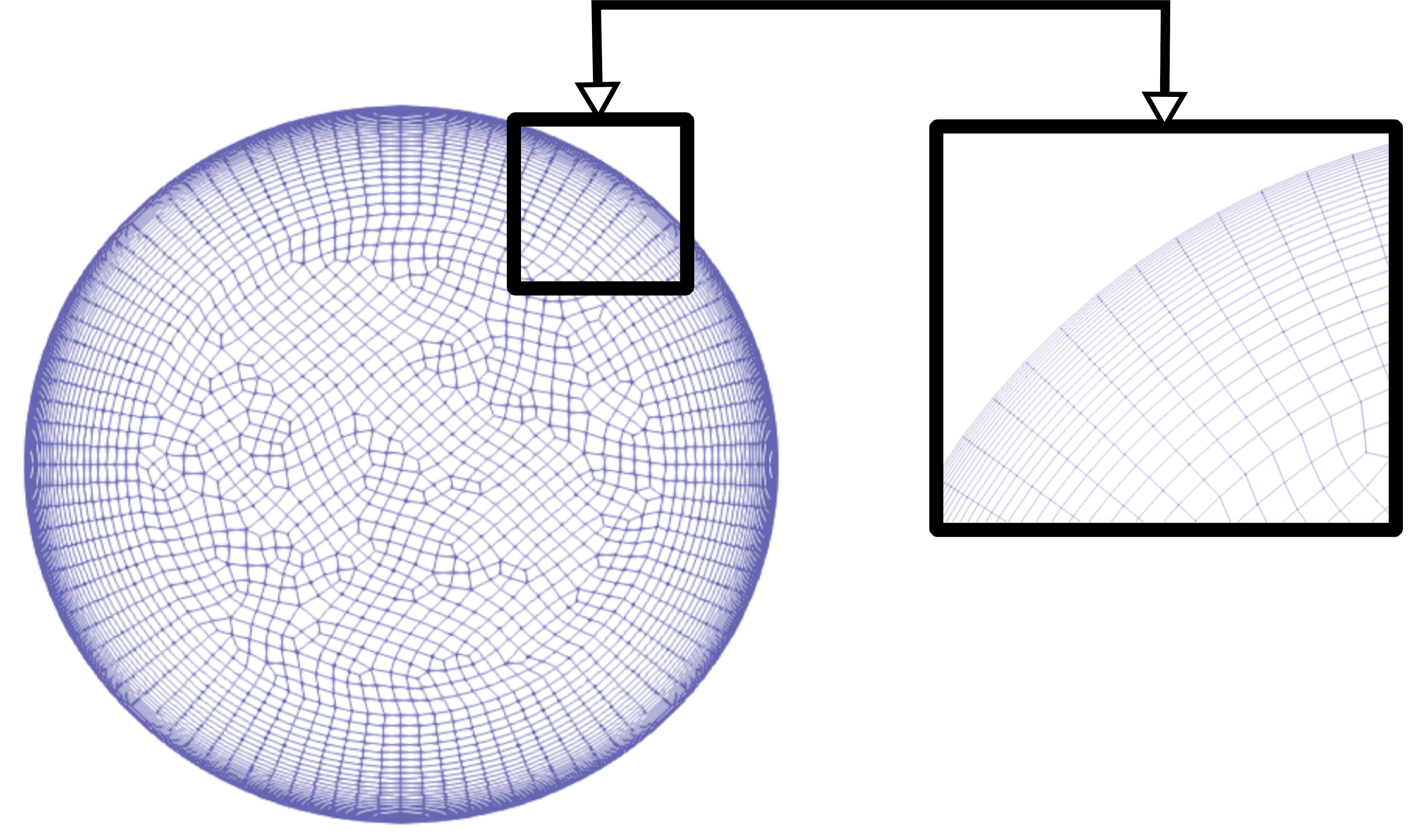}}\\
\caption{(a) Three dimensional view of the computational domain and boundary conditions, (b) detail of the discretization of the pipe section and local magnification of the mesh close to the near-wall cells.}
\label{fig:Computational_domain}
\end{figure}

The turbulent viscosity is modeled using the $k-\omega$ SST model \citep{wilcox1988reassessment} in order to guarantee mixture's pseudo-homogeneity. 
For large scale slurry flows, the $k-\omega$ SST turbulence model should be natural as it properly predicts turbulent variables in the vicinity of walls thanks to the $k-\omega$ model and it allows to achieve turbulent regime in the bulk is achieved by means of $k-\varepsilon$ model.
It is worth noting that the $k-\omega$ SST turbulence model is employed with an enhanced wall function treatment that blends laminar and turbulent laws-of-the-wall to provide continuous and asymptotically correct values for all $y^{+}$. 
A mesh refinement is performed in the near-wall regions.
\textit{A priori} estimate of shear stresses are used to determine the wall-adjacent mesh cell heights required to obtain $y^{+}\approx 1$ (cf. Fig.~\ref{fig:Section_mesh}), where $y^{+}={u}^{\ast}x_2/\nu$ is a function of the friction velocity ${u}^{\ast}=\sqrt{\tau_{\mathrm{wall}}/\rho}$.
The wall shear stress $\tau_{\mathrm{wall}}$ is estimated with the friction factor approximated by the \citet{swamee1976explicit}'s equation and considering the root-mean-squared value of the axial and circumferential Reynolds numbers with zero wall roughness and the average axial velocity.
To enable an effective capturing of the viscous sublayer, the value of $y^{+}$ is kept less than 1.
The growth rate of the mesh is set at \(1.1\) until it reaches the log-law layer, equivalent to \(y^{+}=60\). 
A total of $700 \times10^3$ grid meshes are used to discretize the whole three-dimensional computational domain, as illustrated in Fig.~\ref{fig:3D domain}.
The solver has been able to predict particle profiles for very concentrated sand-water mixtures \citep{elkarii2020towards}.
Computational meshes are generated using \openfoam's \texttt{blockMesh} utility.
The simulations are performed utilizing an implicit-Euler scheme to discretize in time; 
Gradient terms are discretized using a cell limited linear Gauss scheme;
The second order corrected linear Gauss scheme is employed for Laplacian terms;
The interpolation scheme is linear;
For the surface-normal gradient scheme, a corrected scheme is used, and for the wall distance computation, a mesh-wave is used.
Table \ref{schemes} summarizes the major solvers employed in the generation of the present database. 
The reader is referred to \citep{elkarii2022cfd} for more details of the CFD modeling approach and numerical validation of the CFD model.
\begin{table}[ht!]
\centering
\caption{Solver scheme used in the simulations}
\begin{tabular}{c|c}
\hline\hline
Term     & \texttt{TwoPhaseEulerFoam} scheme\\
\hline\hline
Convection \& Momentum transport   & Gauss Van Leer \\
Turbulence \& Artificial compression  & Gauss upwind\\
\hline\hline 
\end{tabular}
\label{schemes}
\end{table}
%
\section{Numerical results}
\label{POD-results}
%
This section assesses the computational efficiency of the proposed POD approach for analyzing slurry flows.
For each test case, we compare the results of the mean and variance of the potential solution using two methods: MC and POD with POD-PCE.
The two-dimensional plan considered in the present study is the mid-plane of the pipe that is a squared domain $\Omega \equiv L_{x_1}\times L_{x_2}= [0, 140D] \times [-0.5D, 0.5D]$.
Consequently, we extract  the two-dimensional two-phase flow solution by employing the deterministic solution obtained from the three-dimensional two-phase Navier-Stokes equations.
In this context, our objective is to evaluate the uncertainty associated with the QoIs, which include the solid concentration, velocity, and pressure distribution over the mid-plane pipe.
%
\subsection{Two-dimensional solid concentration}
\label{2D_C}
%
Initially, we executed 1000 three-dimensional CFD simulations based on a designed experiment, where input parameters were identified and sampled using Latin hypercube sampling to ensure uniform coverage across the designated parameter space for each input.
Subsequently, we obtained the solution on a structured mesh consisting of a total of $18894$ nodes by extracting a mid-plane section.
Next, surrogate models were constructed using POD-PCE methodology.
It is important to highlight that, unlike the conventional PCE method applied to the complete model, the POD-PCE method solves a reduced model.
The accuracy of the POD-PCE method for each eigenvalue is evaluated and presented in Table \ref{POD_modes_C}.
\begin{table}[ht!]
\centering
\caption{Optimal polynomial degrees with LOO errors for the POD modes in the slurry particles concentration ${\phi_{s}}$}
\begin{tabular}{c|c|c}
\hline\hline
Number of modes & LOO error & Polynomial degree\\ 
\hline
Mode \#1    & \(1.2758\times10^{-5}\) & \(6\)  \\
Mode \#2    & \(5.7266\times10^{-4}\) & \(8\)  \\
Mode \#3    & \(3.1746\times10^{-3}\) & \(7\)  \\ 
Mode \#4    & \(1.0955\times10^{-2}\) & \(10\) \\
Mode \#5    & \(3.3131\times10^{-2}\) & \(10\) \\ 
Mode \#6    & \(1.3435\times10^{-1}\) & \(10\) \\
Mode \#7    & \(2.9421\times10^{-1}\) & \(10\) \\ 
Mode \#8    & \(1.0350\times10^{-1}\) & \(10\) \\
Mode \#9    & \(4.7471\times10^{-1}\) & \(10\) \\ 
Mode \#10   & \(7.8437\times10^{-2}\) & \(9\)  \\
Mode \#11   & \(6.7515\times10^{-1}\) & \(10\) \\ 
Mode \#12   & \(1.1624\times10^{-1}\) & \(10\) \\
\hline\hline
\end{tabular}
\label{POD_modes_C}
\end{table}

\begin{figure}[ht!]
\centering
\subfloat[]{\includegraphics[width=0.9\textwidth]{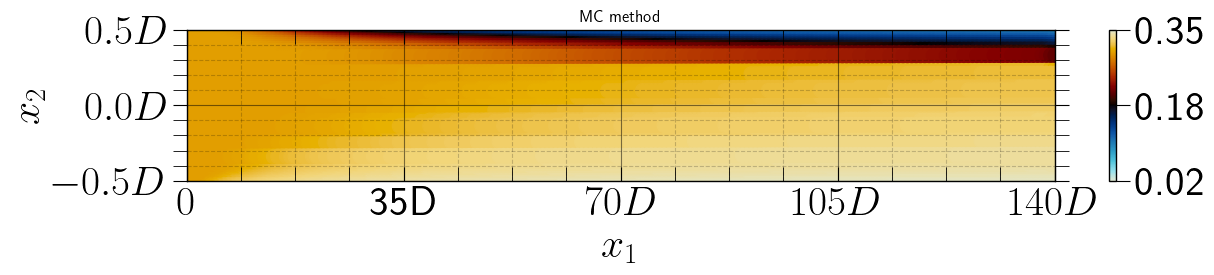}}\\
\subfloat[]{\includegraphics[width=0.9\textwidth]{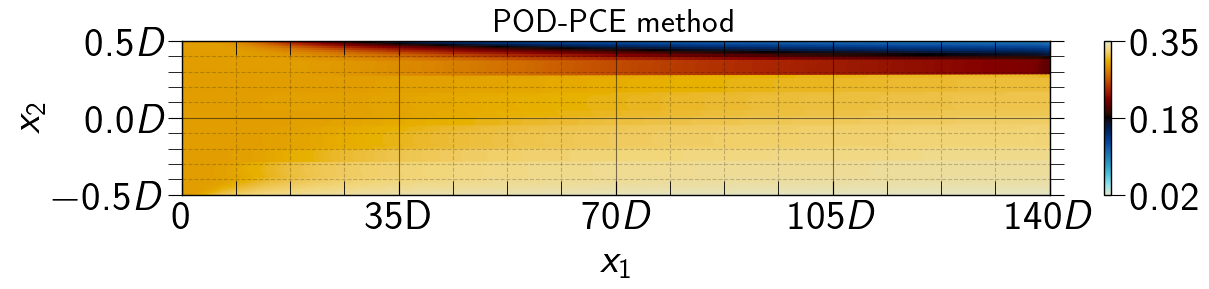}}\\
\subfloat[\label{fig:L2_MEAN_C}]{\includegraphics[width=0.9\textwidth]{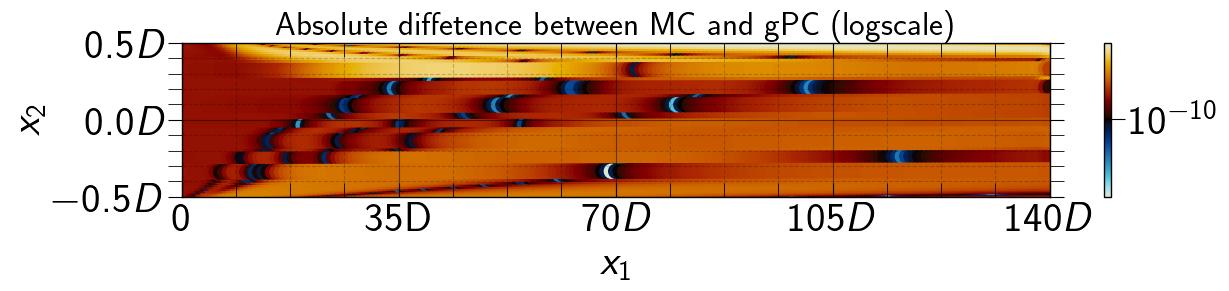}}\\
\caption{Mean solid concentration $\phi_{s}$ obtained for the stochastic simulation (a), the deterministic exact solution (b) and the difference between the two solutions (c) obtained for a slurry flow problem.}
\label{POD_mean_C}
\end{figure}

\begin{figure}[ht!]
\centering
\subfloat[]{\includegraphics[width=0.9\textwidth]{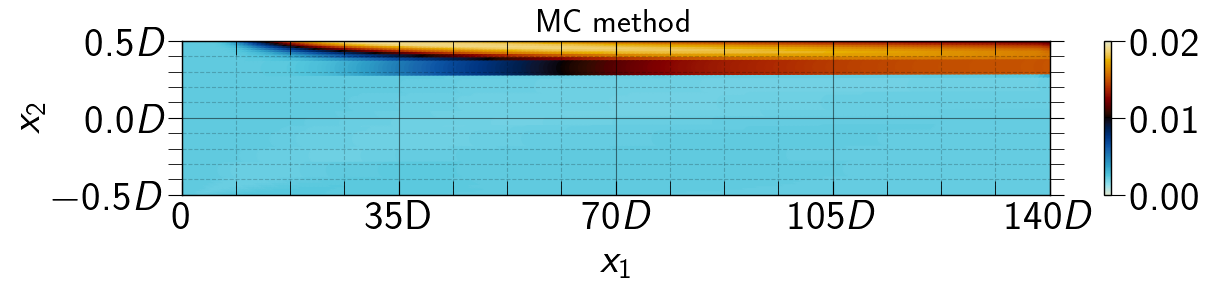}}\\
\subfloat[]{\includegraphics[width=0.9\textwidth]{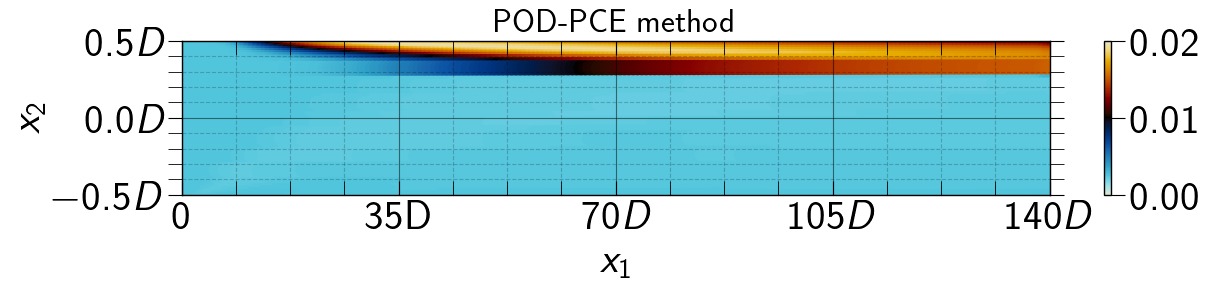}}\\
\subfloat[\label{fig:L2_VAR_C}]{\includegraphics[width=0.9\textwidth]{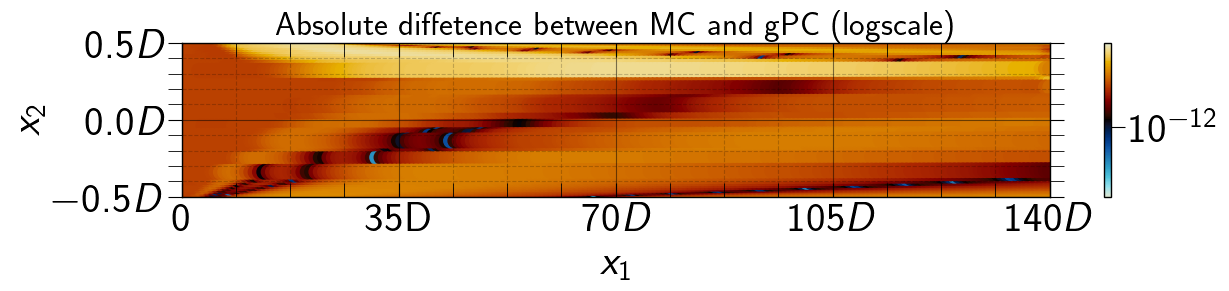}}\\
\caption{Same as Fig.~\ref{POD_mean_C} but for the variance.}
\label{POD_var_C}
\end{figure}

\begin{figure}[ht!]
\centering
\includegraphics[width=0.6\textwidth]{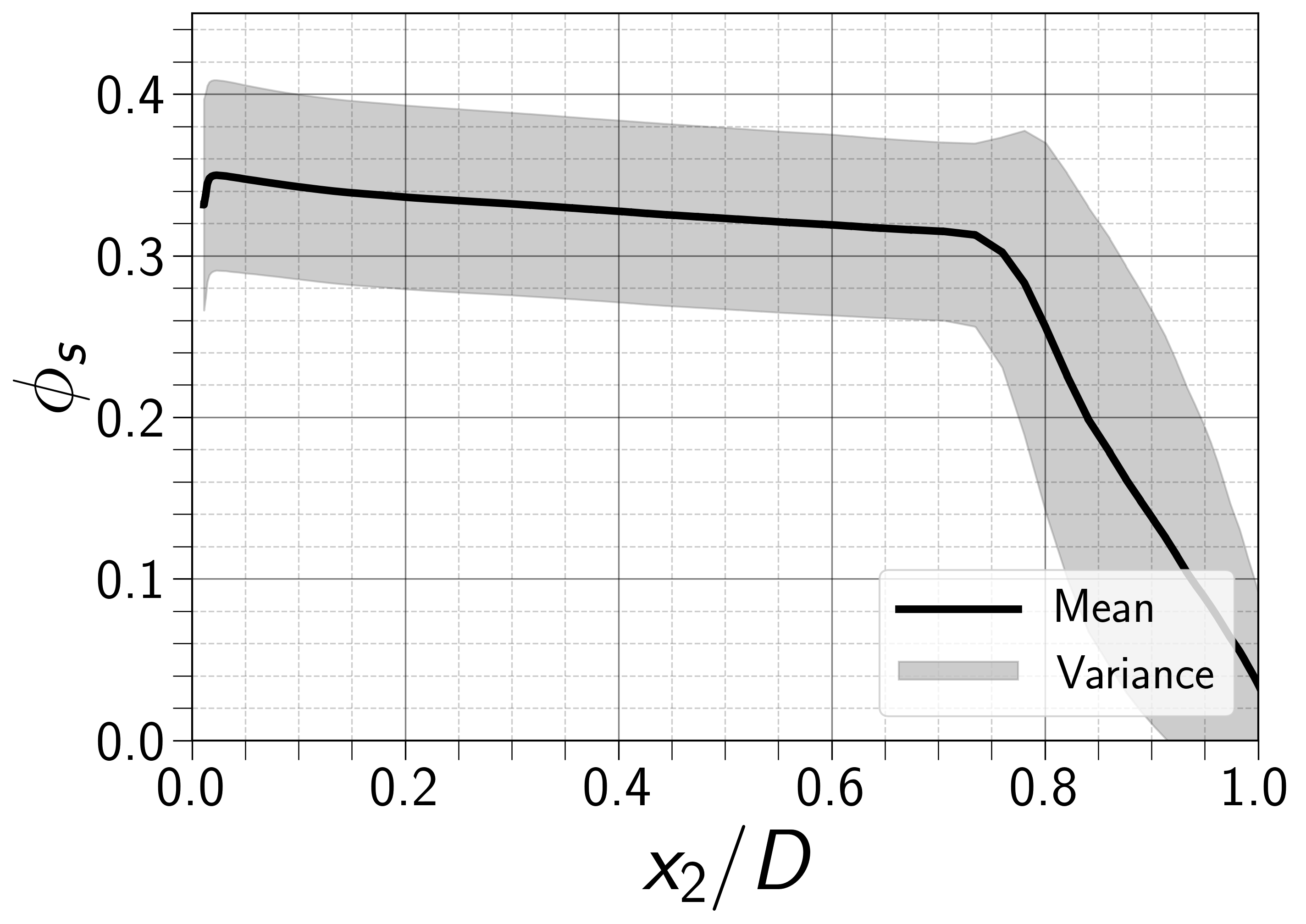}
\caption{Cross-sections of the mean and variance solutions of the solid concentration distribution using PCE-POD}
\label{fig:mean_std_2D_phi}
\end{figure}

In this particular scenario, employing a truncation criterion of $\varepsilon=10^{-4}$, the model selectively retains only the initial $12$ eigenvalues, and the PCE process is carried out individually for each mode.
As a result, the surrogate model significantly reduces computational requirements, necessitating only $12$ decompositions instead of the original $18894$ associated with the total number of nodes in the two-dimensional numerical model.
Table \ref{POD_modes_C} showcases the optimal polynomial degrees employed to approximate each spectral mode using the POD technique, alongside an evaluation of the Leave-One-Out (LOO) error generated by the PCE method when estimating the numerical solution.
Upon analyzing the outcomes, it is evident that the problem exhibits nonlinearity, as high polynomial degrees are required to effectively capture the uncertainty.
Furthermore, the findings indicate satisfactory magnitudes for the LOO errors.
Moving forward, in Fig.~\ref{POD_mean_C}, we compare the mean solid concentration solutions obtained from the stochastic simulation with the deterministic exact solution, while also presenting the discrepancy between the two solutions for easy visual comparison.
The results depicted in Fig.~\ref{POD_mean_C} illustrate that, considering the level of uncertainty examined in the problem, the mean solutions computed by both methods exhibit similar trends and patterns.
The stochastic approach successfully addresses the distribution of solid concentration, producing numerical outcomes that are devoid of any nonphysical oscillations.
In general, these findings imply that the model reduction has minimal impact on the accuracy of UQ.
Furthermore, Fig.~\ref{POD_var_C} showcases the obtained results for the variance.
The results for mean and variance make it evident that the two methods under consideration exhibit identical flow patterns and accurately capture the solid concentration over the two-dimensional pipe.
To provide a more comprehensive understanding, Fig~\ref{fig:L2_MEAN_C} and Fig.~\ref{fig:L2_VAR_C} illustrate the logarithmic $\ell_2$-error in estimating the statistical moments considered.
The remarkably low magnitude of the error $(< 10^{-10})$ clearly indicates that the surrogate models proficiently capture the uncertainty manifested over the physical domain for the slurry flow problem under consideration.
An additional benefit is that the proposed POD-PCE method attains this high level of accuracy while incurring significantly lower computational costs in comparison to the MC and PCE methods.
For a better insight, we display in Fig.~\ref{fig:mean_std_2D_phi} vertical cross-sections of the mean and variance at $x_1=126D$.

The distribution of the mean solid concentration reflects the general behavior of the flow in the case of phosphate slurry.
The findings indicating that the average concentration of particles in the phosphate slurry flow is stratified into two distinct layers. 
This confirms that the slurry flow is of the settling type.
In the core and lower wall of the pipe where particles move slowly, the solid concentration variance is insignificant. 
However, the top wall exhibits the highest variance values due to the relatively sparse flow and increased particle movement and kinetic energy.

\begin{figure}[ht!]
\centering
\includegraphics[width=0.6\textwidth]{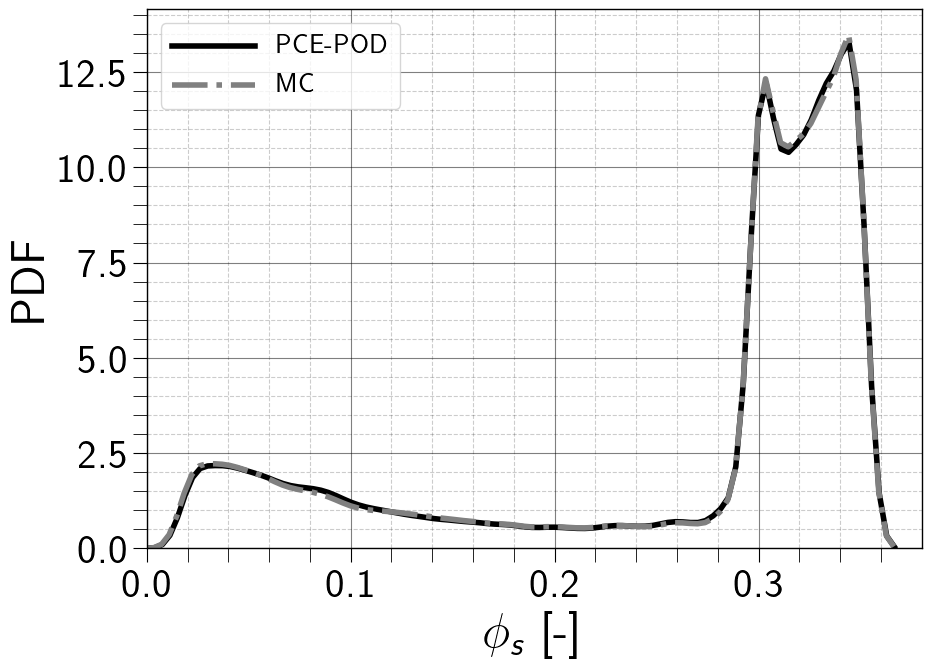}
\caption{Visualization of solid concentration $(\phi_{s})$ PDF through Kernel smoothing (KS) method, estimated using MC and PCE-POD techniques.}
\label{fig:KS_POD_C}
\end{figure}

The ability of the surrogate model to estimate the whole Probability Density Function (PDF) of the solids concentration is also investigated.
This comparison is shown in Fig.~\ref{fig:KS_POD_C} where the estimation of the whole PDF using Monte-Carlo simulation are compared to those obtained using the PCE-POD surrogate model.  PDF estimation is done here using a non-parametric estimation following a kernel smoothing methodology.
It is clear that the results demonstrate that there is a good estimation of the probability density function using the tools presented in this study.
The results demonstrate a robust convergence in the decomposition, rendering them suitable for UQ in the distribution of solid concentration in slurry flows.
The general shape of the PDF exhibits also the non-linearity effects, as it is not suited to a Gaussian-like shape.
One can also see the presence of two modes.
This may be due to segregation effects.
The slurry flow experiences segregation phenomena during transportation and the PDF of the slide concentration indicates the formation of two distinct modes in the concentration distribution.

\begin{figure}[ht!]
\centering
\subfloat[]{\includegraphics[width=0.45\textwidth]{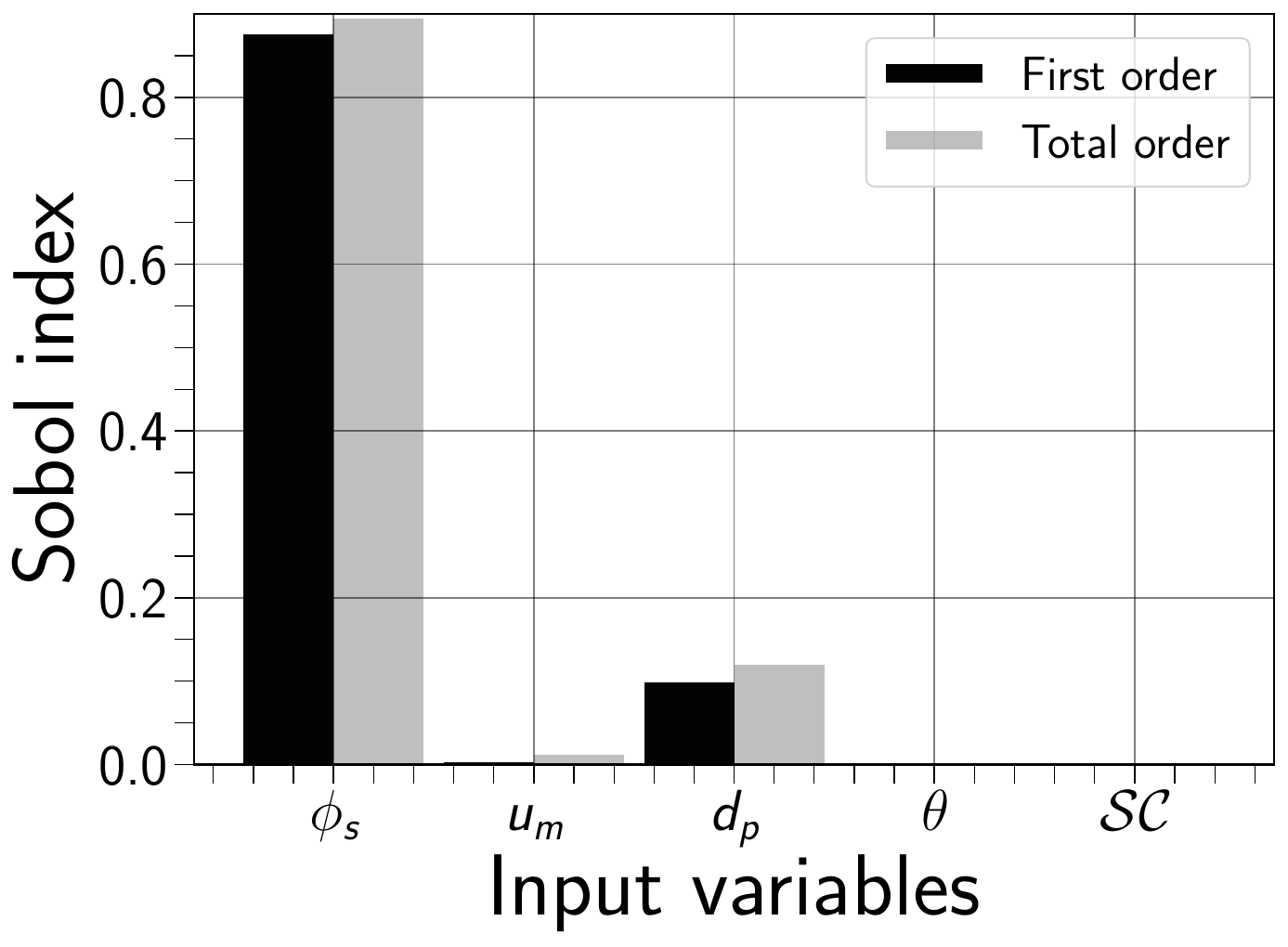}}\hfill
\subfloat[]{\includegraphics[width=0.45\textwidth]{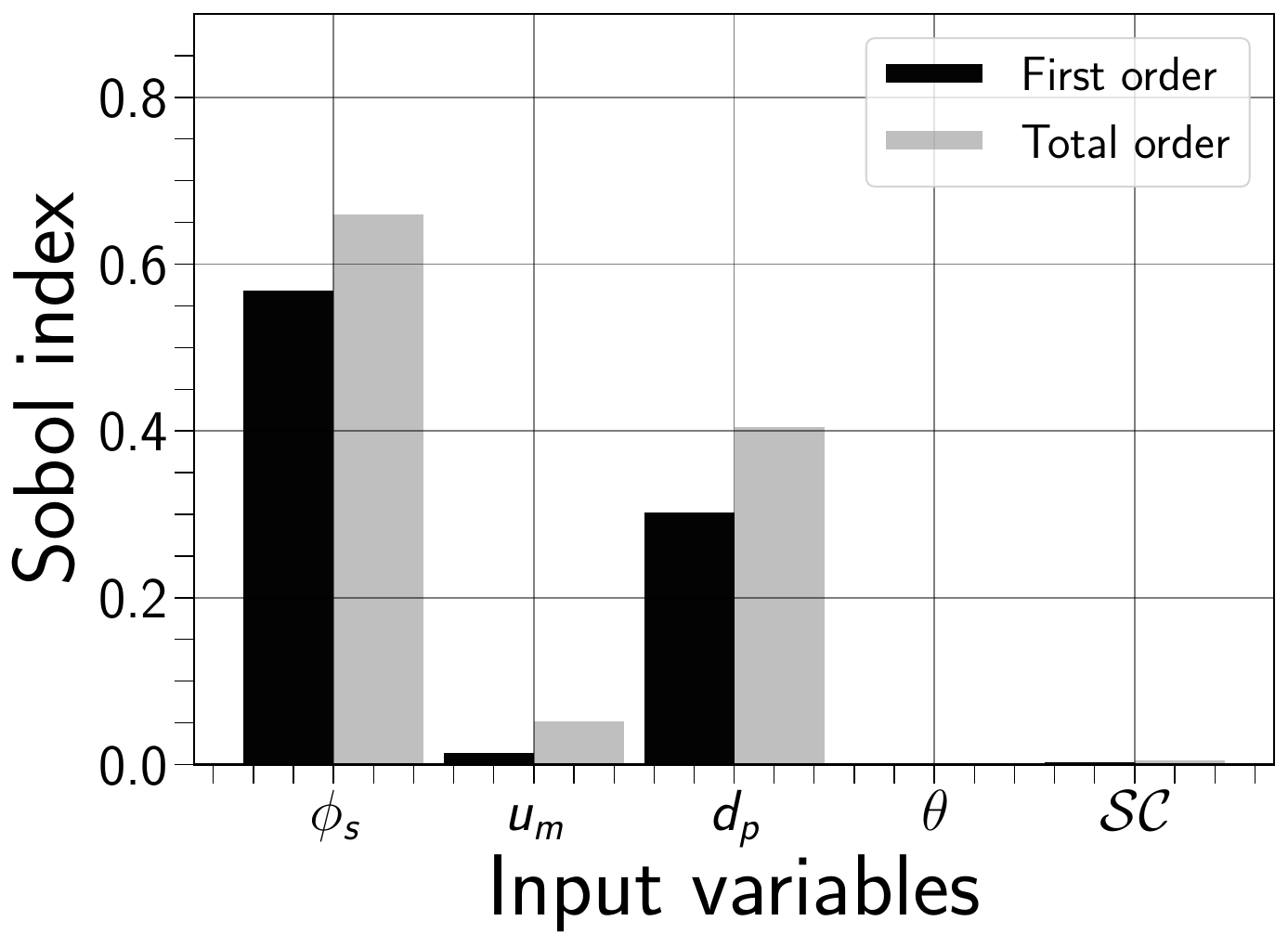}}\\
\caption{Barplots of first and total Sobol’ indices for solid concentration field calculated based on: (a) One-dimensional data, (b) Two-dimensional data }
\label{fig:2D_C_agregé}
\end{figure}

The sensitivity of the solid concentration field to the variability of the random parameters $\left\{\phi_{s}, u_{m}, d_{p} , \theta, \mathcal{SC}\right\}$, can be assessed by means of bar-plots of the first order and total aggregated Sobol' indices (cf. Fig.~\ref{fig:2D_C_agregé}).
It is important to mention that a sensitivity analysis using only the PCE method has already been conducted for one-dimensional and
scalar QoIs \citep{elkarii2023global}.
The corresponding Sobol' indices results will be presented alongside the outcomes obtained through the POD-PCE method, with the intention of facilitating a comprehensive comparison.
The objective of this comparison is to assess whether increasing the dimensionality of our QoIs has any impact on the obtained results.
The reduced model generates a comparable ranking of the most influential parameters; however, there are variations in the magnitude of the Sobol' indices. 
The Sobol' indices analysis conducted indicates that the distribution of solid concentration along the vertical diameter of the pipe remains primarily influenced by the initial concentration introduced into the pipeline, followed by the particle sizes.
Thus, these two parameters are responsible for regulating the behavior and pattern of the slurry flow.
Indeed, the inlet concentration of a slurry directly impacts the mixture density, and it plays a significant role in the behavior of particles, including their collisions and friction, ultimately leading to their arrangement inside the pipeline.
The behavior of particles is also induced by their size, as larger particles generally cause sedimentation, resulting in a significant change in the in situ solid concentration in the pipeline.
%
\subsection{Two-dimensional solid velocity}
%
Our second QoI consists of a two-dimensional velocity field of particles.
Similar to the previous example, we have provided the optimal polynomial degree distribution for approximating each spectral mode using the POD technique (cf. Table \ref{POD_modes_u}). 
Additionally, we have included the Leave-One-Out (LOO) error associated with the PCE method for estimating the numerical solution for each mode.
\begin{table}[ht!]
\centering
\caption{Optimal polynomial degrees with LOO errors for the POD modes in the slurry particles velocity ${u_{s}}$}
\begin{tabular}{cccccc}\hline\hline
Number of modes & LOO error & Polynomial degree \\ 
\hline\hline
Mode \#1 & \(4.4253\times 10^{-5}\) & \(7\) \\ 
Mode \#2 & \(2.0990\times 10^{-3}\) & \(9\) \\
Mode \#3 & \(1.1500\times 10^{-2}\) & \(6\) \\ 
Mode \#4 & \(2.0464\times 10^{-2}\) & \(7\) \\
Mode \#5 & \(3.9424\times 10^{-2}\) & \(8\) \\
\hline\hline
\end{tabular}
\label{POD_modes_u}
\end{table}
By using a set of $5$ eigenmodes, the physical model can be significantly reduced.
This means that the surrogate model for two-dimensional solid velocity can be represented by only the first $5$ modes, rather than the $18894$ required to represent the total number of nodes in the two-dimensional numerical model.

\begin{figure}[ht!]
\centering
\subfloat[]{\includegraphics[width=0.9\textwidth]{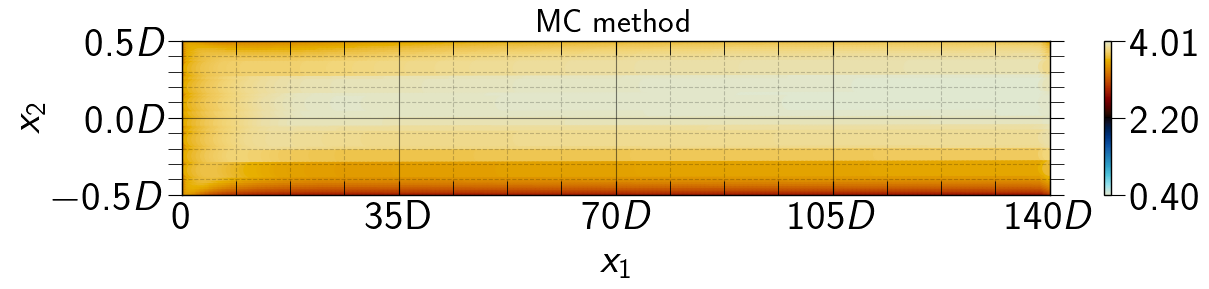}}\\
\subfloat[]{\includegraphics[width=0.9\textwidth]{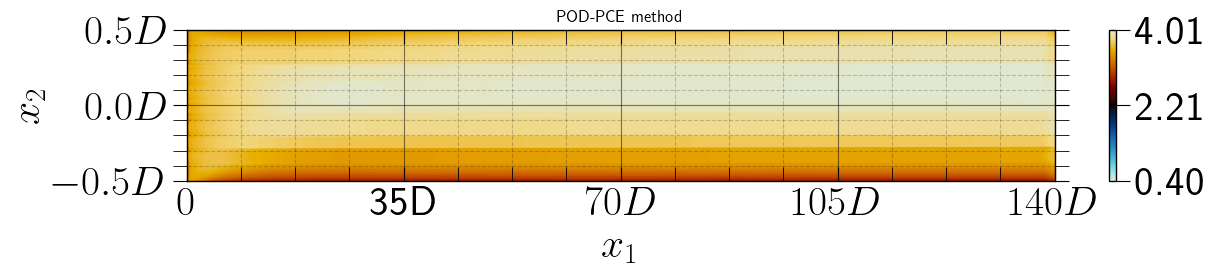}}\\
\subfloat[]{\includegraphics[width=0.9\textwidth]{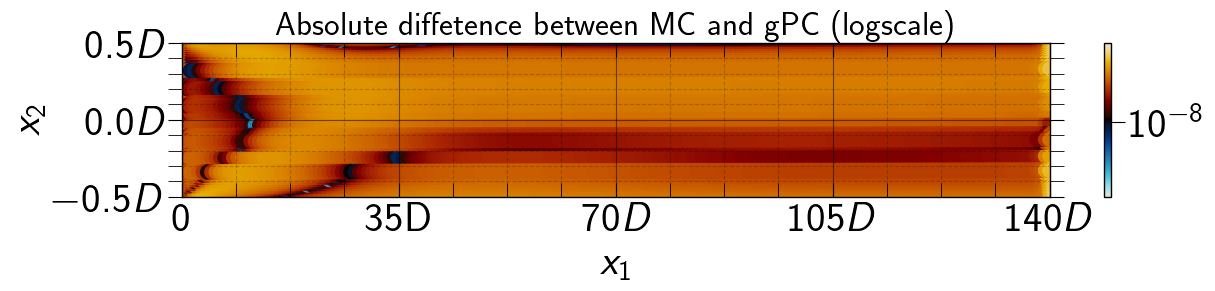}}\\
\caption{Mean solid velocity ${u}_{s}$ obtained for the stochastic simulation (a), the deterministic exact solution (b) and the difference between
the two solutions (c) obtained for a slurry flow problem.}
\label{POD_mean_u}
\end{figure}

\begin{figure}[ht!]
\centering
\subfloat[]{\includegraphics[width=0.9\textwidth]{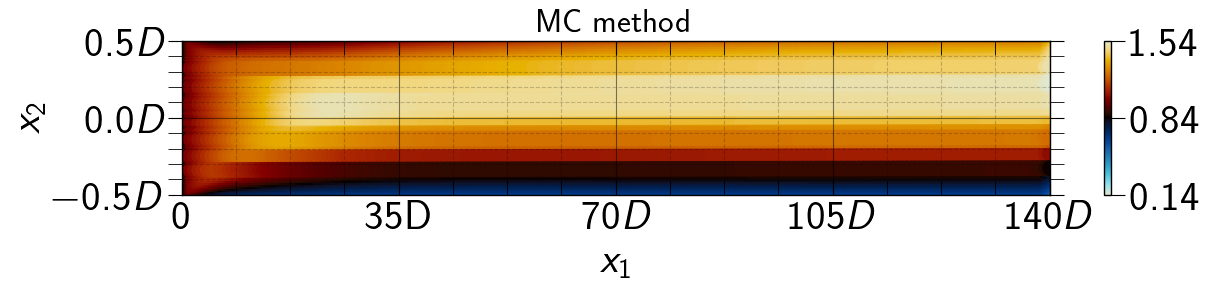}}\\
\subfloat[]{\includegraphics[width=0.9\textwidth]{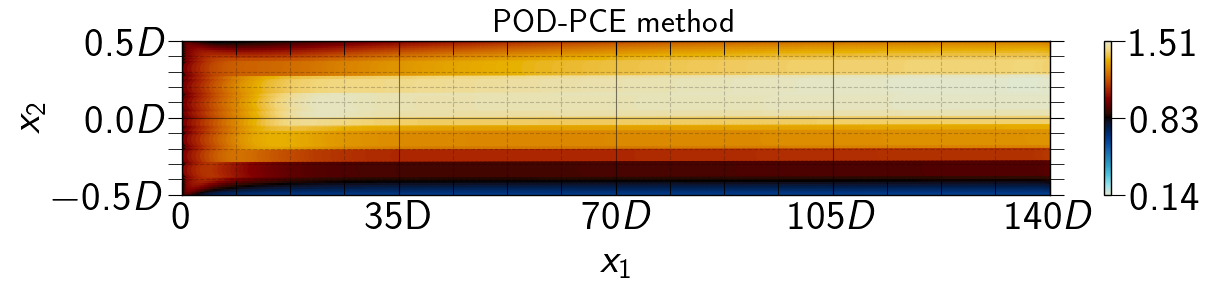}}\\
\subfloat[]{\includegraphics[width=0.9\textwidth]{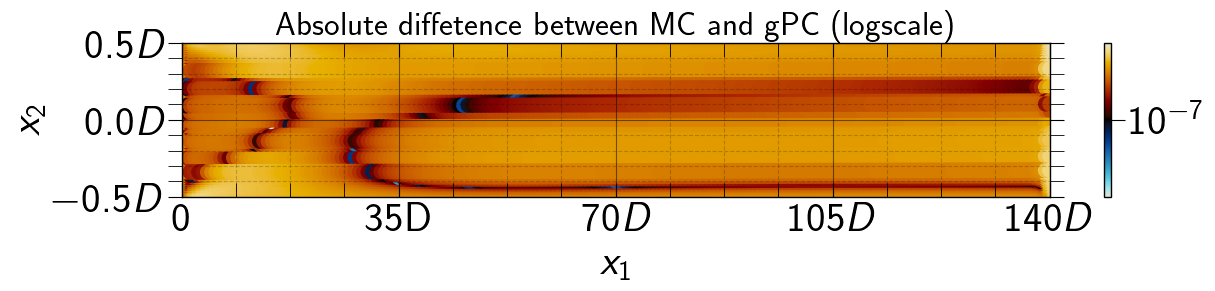}}\\
\caption{Same as Fig.~\ref{POD_mean_u} but for the variance.}
\label{POD_var_u}
\end{figure}

\begin{figure}[ht!]
\centering
\includegraphics[width=0.6\textwidth]{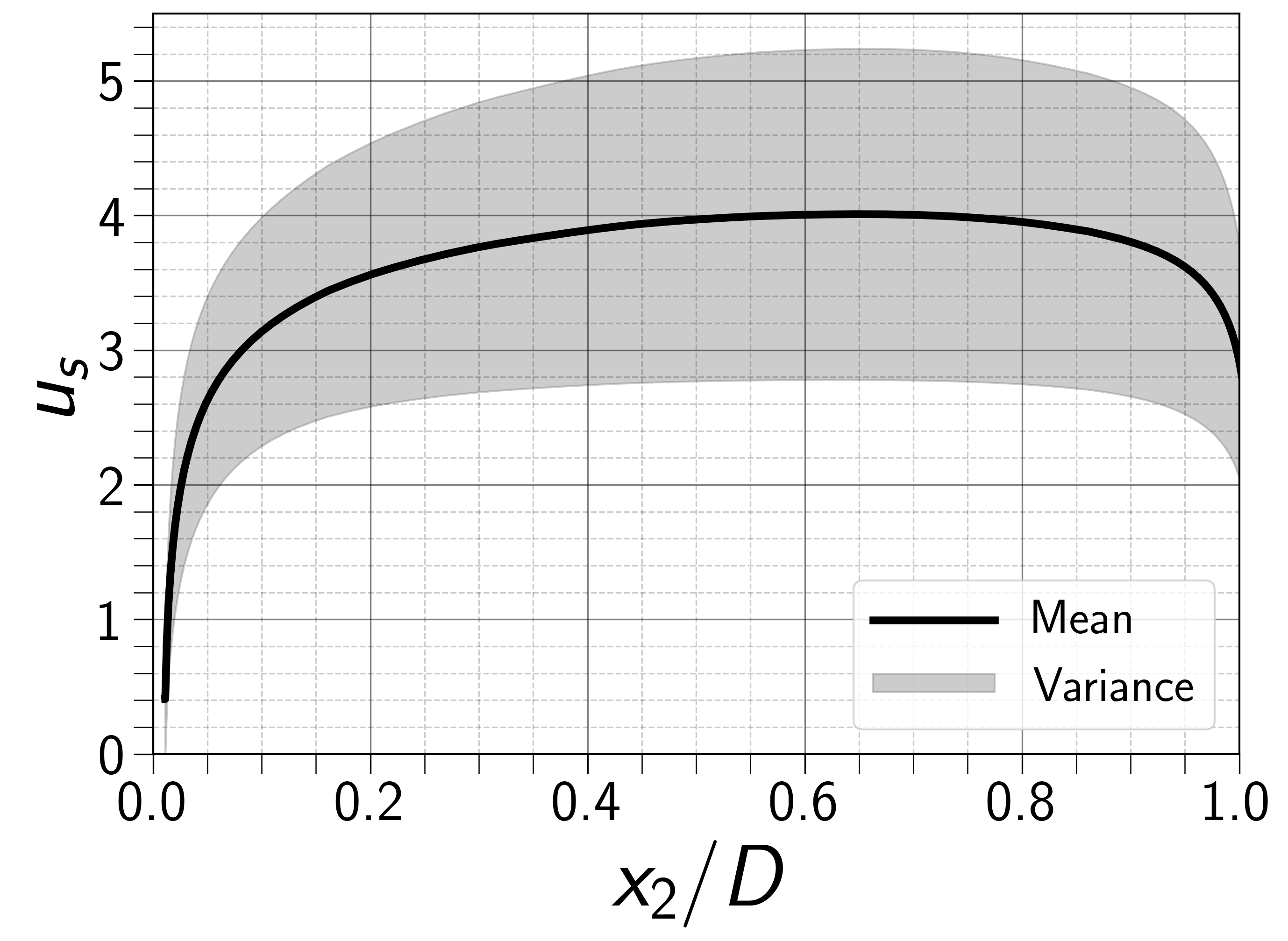}
\caption{Cross-sections of the mean and variance solutions of the solid velocity distribution using PCE-POD}
\label{fig:mean_std_2D_u}
\end{figure}

The ability of the proposed POD-PCE method for recovering the particle velocity of a plane pipe is examined.
We showcase in Fig.~\ref{POD_mean_u} the results estimated using MC and the proposed surrogate model for the mean solid velocity, as well as , the difference between them.
Under the considered conditions of the slurry flow, both methods show comparable trends in the velocity profiles of particles.
Fig.~\ref{POD_var_u} presents the comparative results obtained for the variance.
The magnitude of logarithmic $\ell_2$-error being less than $(< 10^{-6})$ indicates that both methods show comparable velocity profile trends for the solid phase.
Once again, the results that have been presented demonstrate a strong convergence in the decomposition process, making them trustworthy for UQ purposes.
This confirms that the proposed POD-PCE approach is effective in resolving the issue of the two-dimensional velocity distribution of particles.
Fig.~\ref{fig:mean_std_2D_u} depicts the vertical cross-sections of the mean and variance at $x_1 = 126D$.

The presence of an asymmetrical profile in the distribution of mean solid velocity is likely a contributing factor to the observed differences in solid concentration variance across the various walls of the pipe.
Specifically, the skewness of the velocity profile towards the top wall indicates that the flow in that region is relatively more diluted, and the particles have a higher kinetic energy. 
This is consistent with the highest variance values observed in that area.

\begin{figure}[ht!]
\centering
\includegraphics[width=0.6\textwidth]{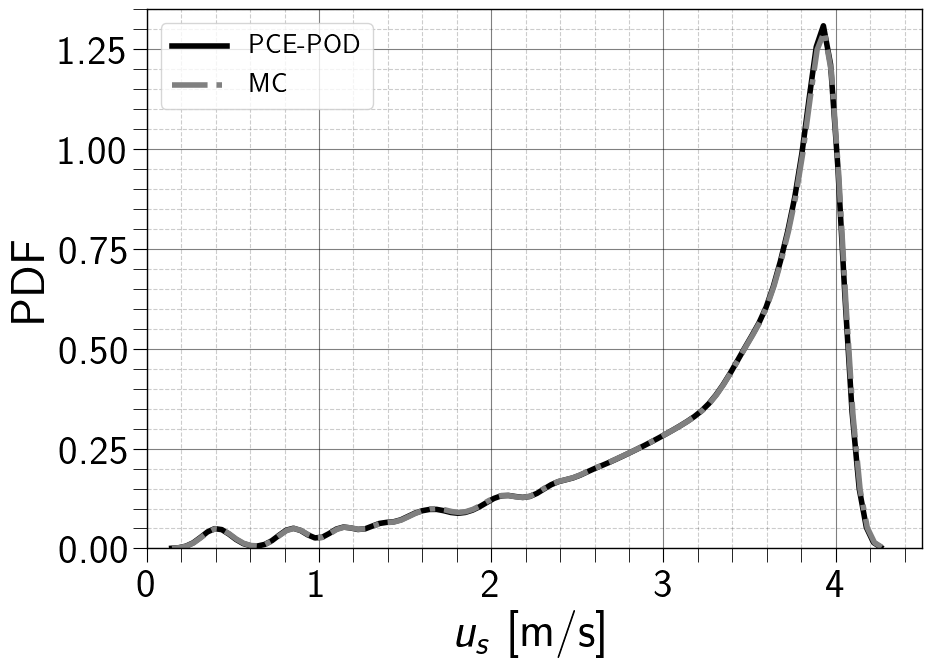}
\caption{Visualization of solids velocity $(u_{s})$ PDF through Kernel smoothing (KS) method, estimated using MC and PCE-POD techniques.}
\label{fig:KS_POD_U}
\end{figure}

The surrogate model ability to estimate the whole Probability Density Function (PDF) of the solids velocity is also checked here.
This comparison is shown in Fig.~\ref{fig:KS_POD_U} where the estimation of the whole PDF using Monte-Carlo simulation are compared to those obtained using the PCE-POD surrogate model. 
Once again, the nonlinear behavior of the model and its impacts on the velocity distribution is emphasized.  This considers a further assessment of the ability of the surrogate model to perform UQ correctly. 

Similar to $\phi_{s}$ the sensitivity of the solid velocity field to the variability of our random parameters can also be evaluated through barplots of the first-order and total Sobol' indices.
\begin{figure}[ht!]
\centering
\subfloat[]{\includegraphics[width=0.45\textwidth]{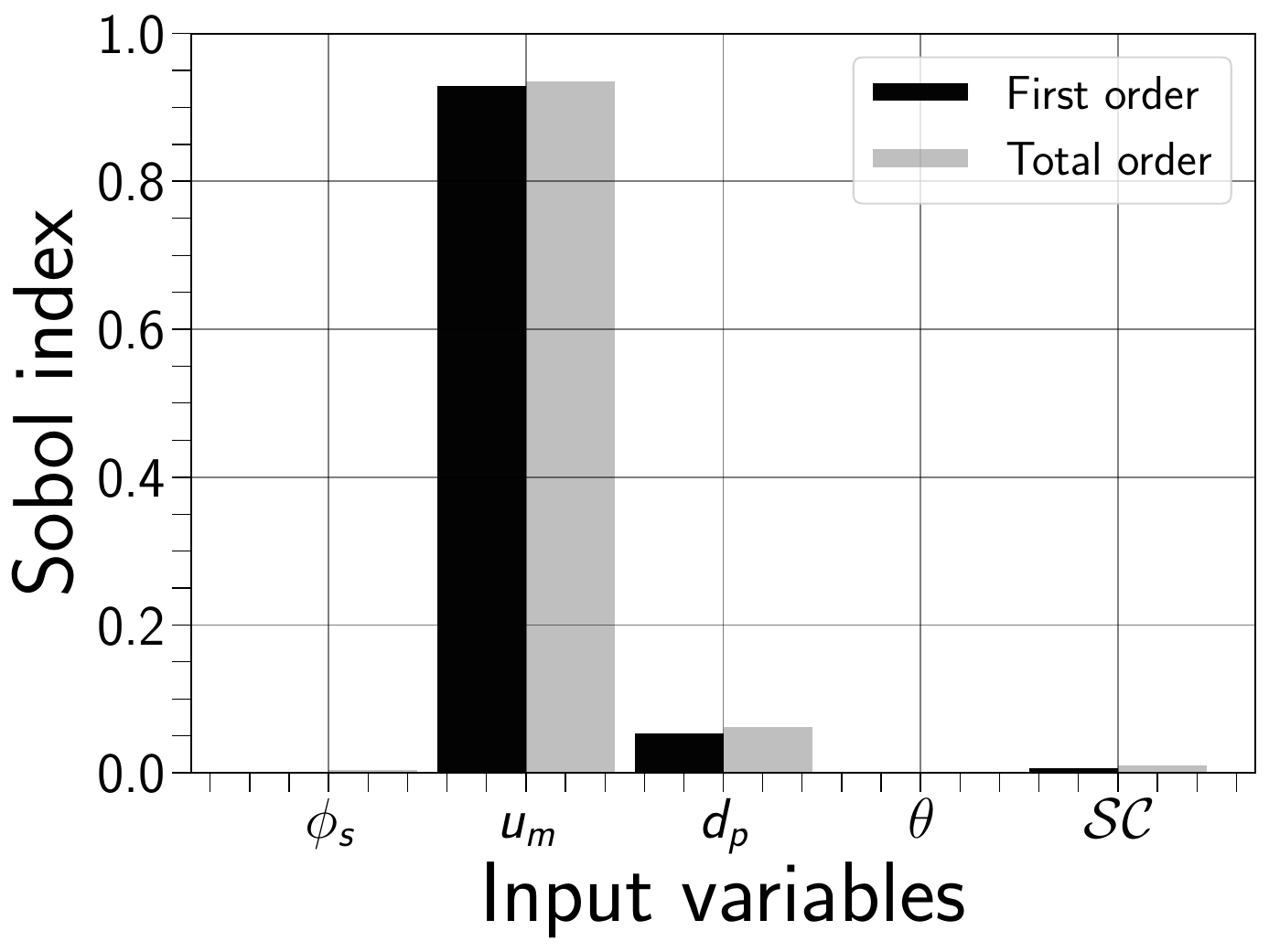}}\hfill
\subfloat[]{\includegraphics[width=0.45\textwidth]{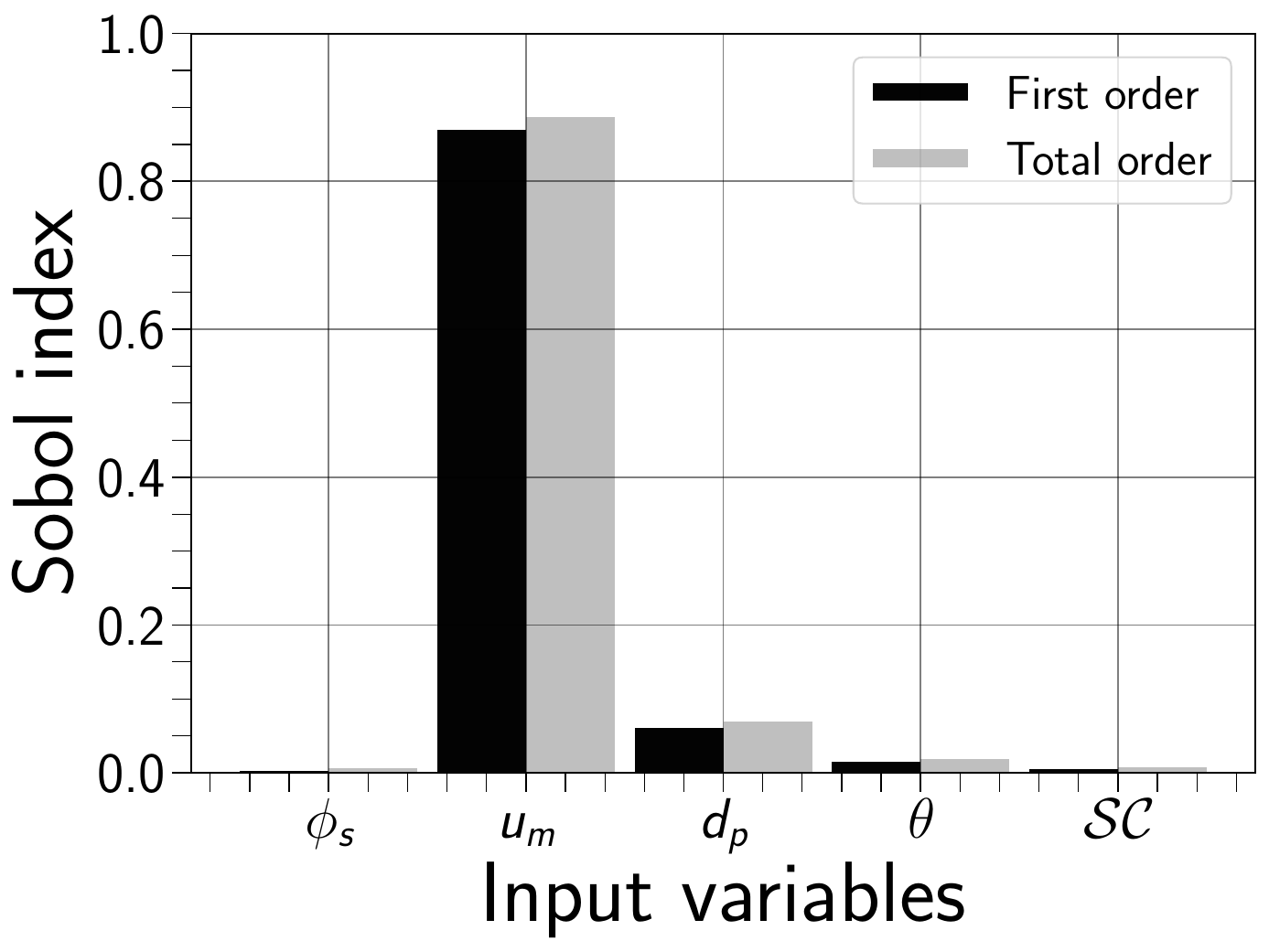}}\\
\caption{Barplot of first and total Sobol’ indices for solid velocity calculated based on: (a) One-dimensional data, (b) Two-dimensional data }
\label{fig:2D_U_agrege}
\end{figure} 
The velocity distribution of solids in the pipe, as depicted in Fig.~\ref{fig:2D_U_agrege}, is primarily influenced by the inlet flow rate. 
On the other hand, the utilization of the POD-PCE technique reveals slightly dissimilar outcomes, indicating that the velocity distribution of particles is relatively more affected by their sizes and also slightly by the pipe inclination.
Based on their values, the observed discrepancies could be attributed to the stochastic nature of the methodology. 
Normally, smaller particles have higher velocities than larger particles at the same temperature.
This is because smaller particles have a higher surface area-to-volume ratio, which means they have more contact with the surrounding fluid, allowing them to exchange energy and momentum more efficiently.
Additionally, smaller particles experience less drag and resistance from the surrounding fluid, which can also contribute to their higher velocity. 
%
%
\subsection{Two-dimensional pressure distribution}
%
The pressure distribution within the slurry flow is the third QoI that was addressed in this study.
Table \ref{POD_modes_p} summarizes the results obtained for PCE-POD modes.
\begin{table}[ht!]
\centering
\caption{Optimal polynomial degrees with LOO errors for the POD modes in the slurry flow pressure $P$}
\begin{tabular}{cccccc}\hline\hline
Number of modes & LOO error & Polynomial degree \\ 
\hline\hline
Mode \#1 & \(6.0655\times 10^{-3}\) & \(6\) \\ 
Mode \#2 & \(3.1614\times 10^{-3}\) & \(6\) \\
\hline\hline
\end{tabular}
\label{POD_modes_p}
\end{table}
It can be observed that the model size has significantly decreased, as the pressure distribution of the slurry flow can be reconstructed using only two modes instead of the $18894$ that were required for obtaining the deterministic solution.

\begin{figure}[ht!]
\centering
\subfloat[]{\includegraphics[width=0.9\textwidth]{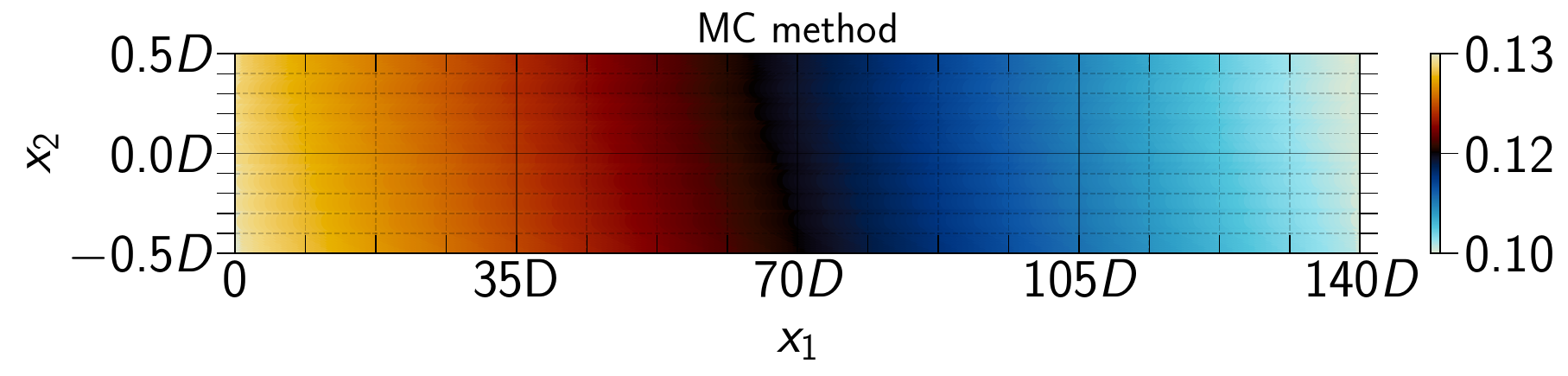}}\\
\subfloat[]{\includegraphics[width=0.9\textwidth]{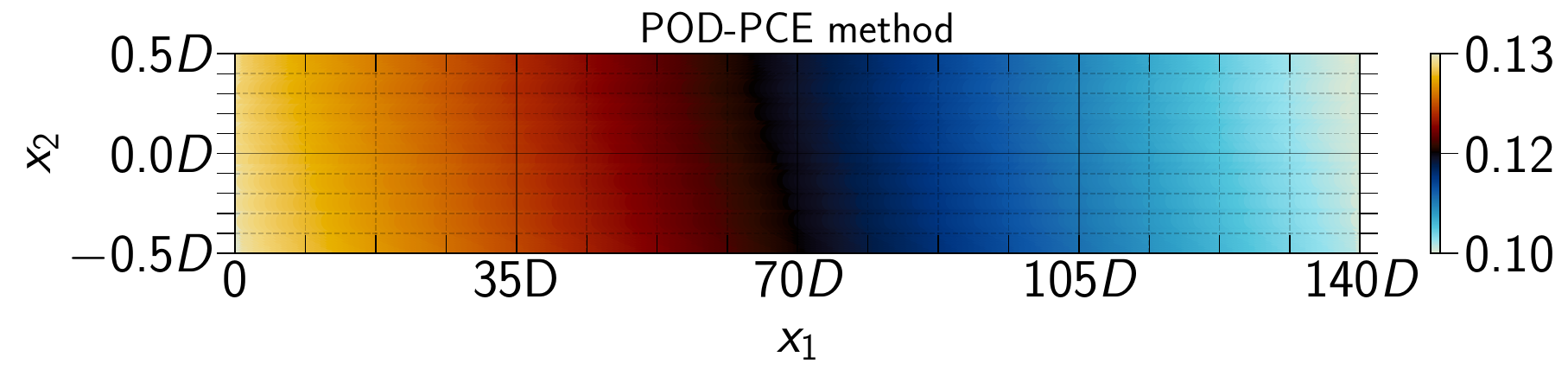}}\\
\subfloat[]{\includegraphics[width=0.9\textwidth]{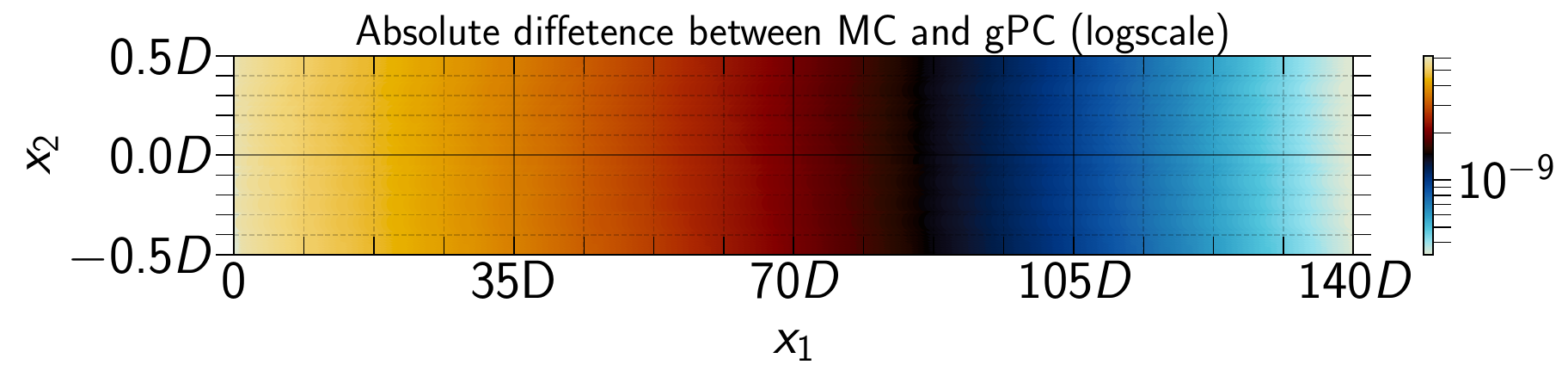}}\\
\caption{Mean flow pressure $P$ obtained for the stochastic simulation (a), the deterministic exact solution (b) and the difference between
the two solutions (c) obtained for a slurry flow problem.}
\label{POD_mean_p}
\end{figure}

Similar to previous analyses, we compare the average pressure distribution obtained from both MC and POD-PCE methods to evaluate the capability of the proposed POD-PCE approach in reconstructing the pressure of the two-dimensional pipe flow.
One should note that the results displayed here for the pressure are calculated in bar unit. The outcomes are depicted in Fig.~\ref{POD_mean_p}.
There are no significant changes between the two solutions, it is further confirmed by the logarithmic $\ell_2$-error estimation. Since, the error's magnitude depicted here is less than $10^{-8}$.

\begin{figure}[ht!]
\centering
\subfloat[]{\includegraphics[width=0.9\textwidth]{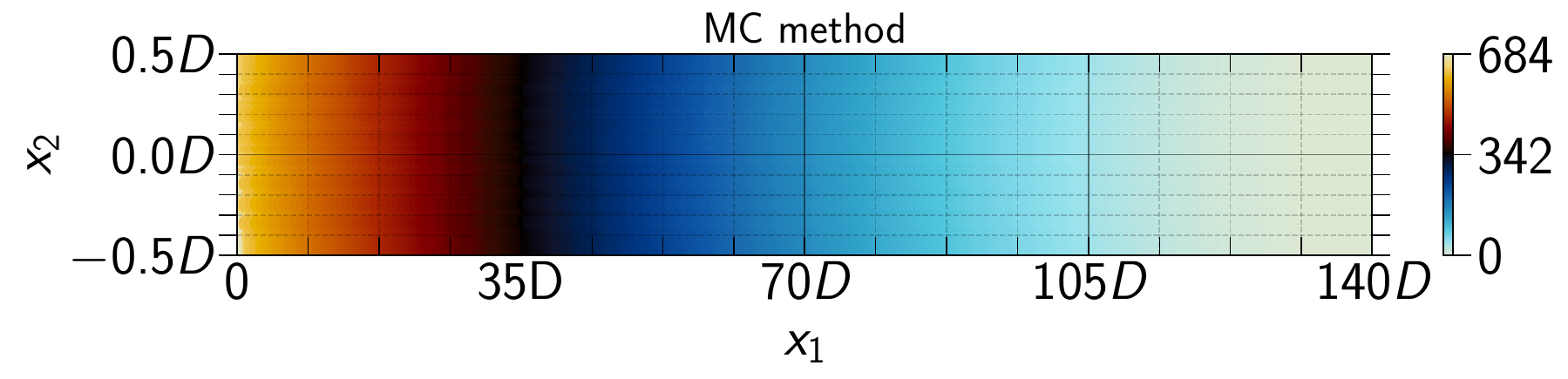}}\\
\subfloat[]{\includegraphics[width=0.9\textwidth]{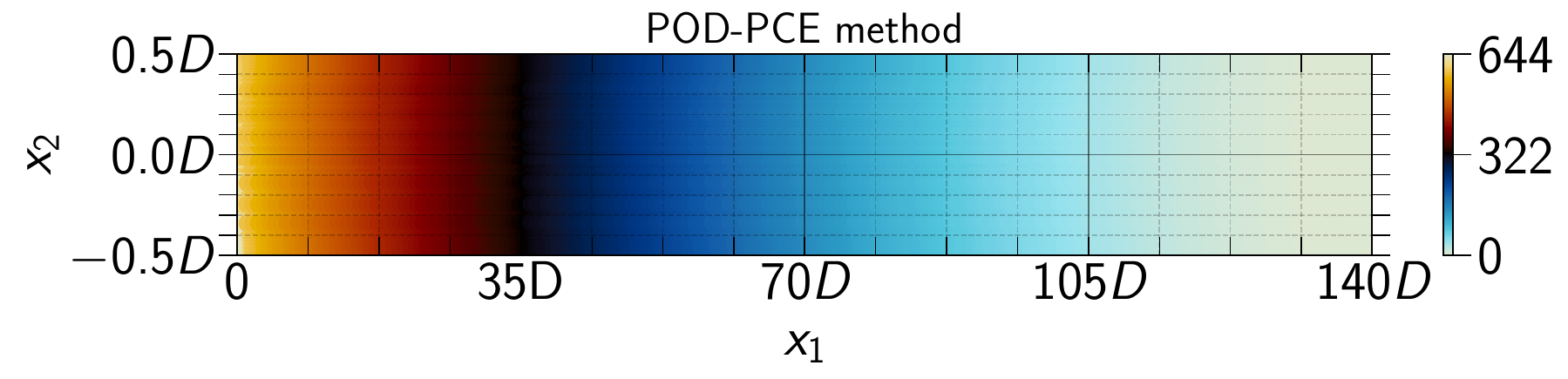}}\\
\subfloat[]{\includegraphics[width=0.9\textwidth]{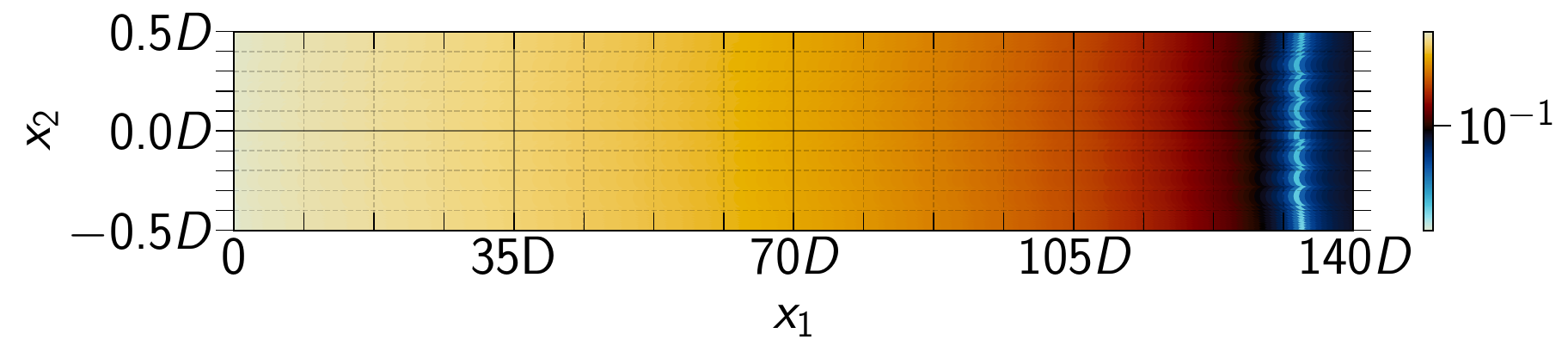}}\\
\caption{Same as Fig.~\ref{POD_mean_p} but for the variance.}
\label{POD_var_p}
\end{figure}

Fig.~\ref{POD_var_p} presents the comparative results obtained for the variance.
The surrogate models effectively capture the uncertainty observed across the physical domain in the slurry flow problem under consideration.
Given the relatively low computational effort required to implement the PCE-POD method, it can be deemed an ideal algorithm for conducting uncertainty quantification in computational slurry flows involving stochastic inputs.
By analyzing the distribution of the mean flow pressure, it can be deduced that the pressure of the slurry typically decreases within the range of variability of the random inputs.
This pressure evolution along the pipeline is evident from the stratified flow pattern discussed in subsection \ref{2D_C}.
Such slurry flows exhibit fluctuations in particle concentration and velocity, resulting in frictional resistance and subsequent pressure decrease.
The highest variance values are observed at the inlet of the pipe because the pressure at the inlet varies in conjunction with the velocity.

\begin{figure}[ht!]
\centering
\includegraphics[width=0.6\textwidth]{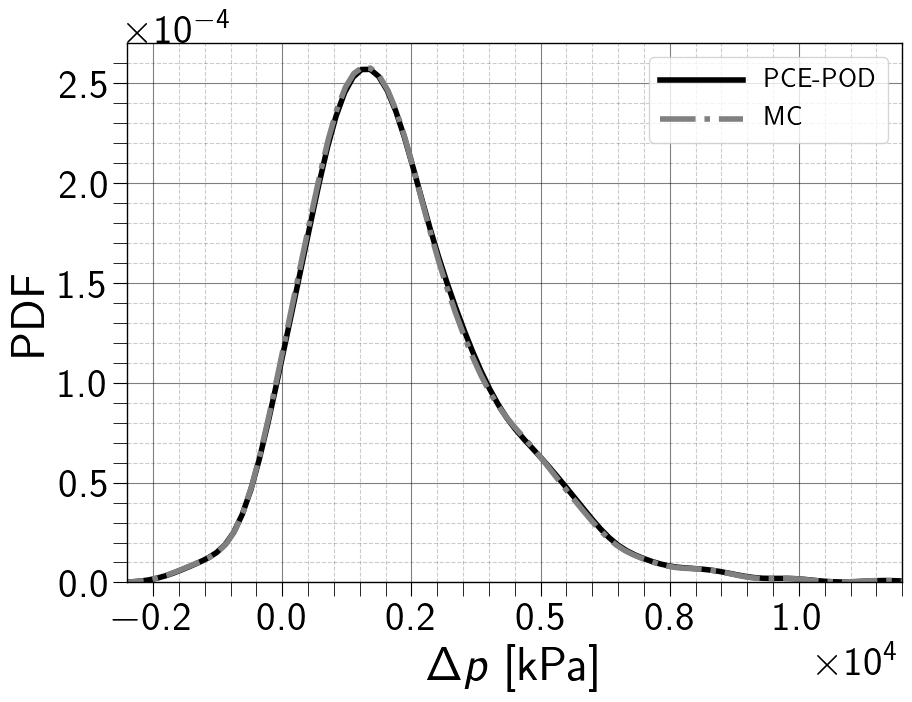}
\caption{Visualization of the flow's pressure drop PDF through Kernel smoothing (KS) method, estimated using MC and PCE-POD techniques.}
\label{fig:KS_POD_DP}
\end{figure}

\begin{figure}[ht!]
\centering
\subfloat[]{\includegraphics[width=0.45\textwidth]{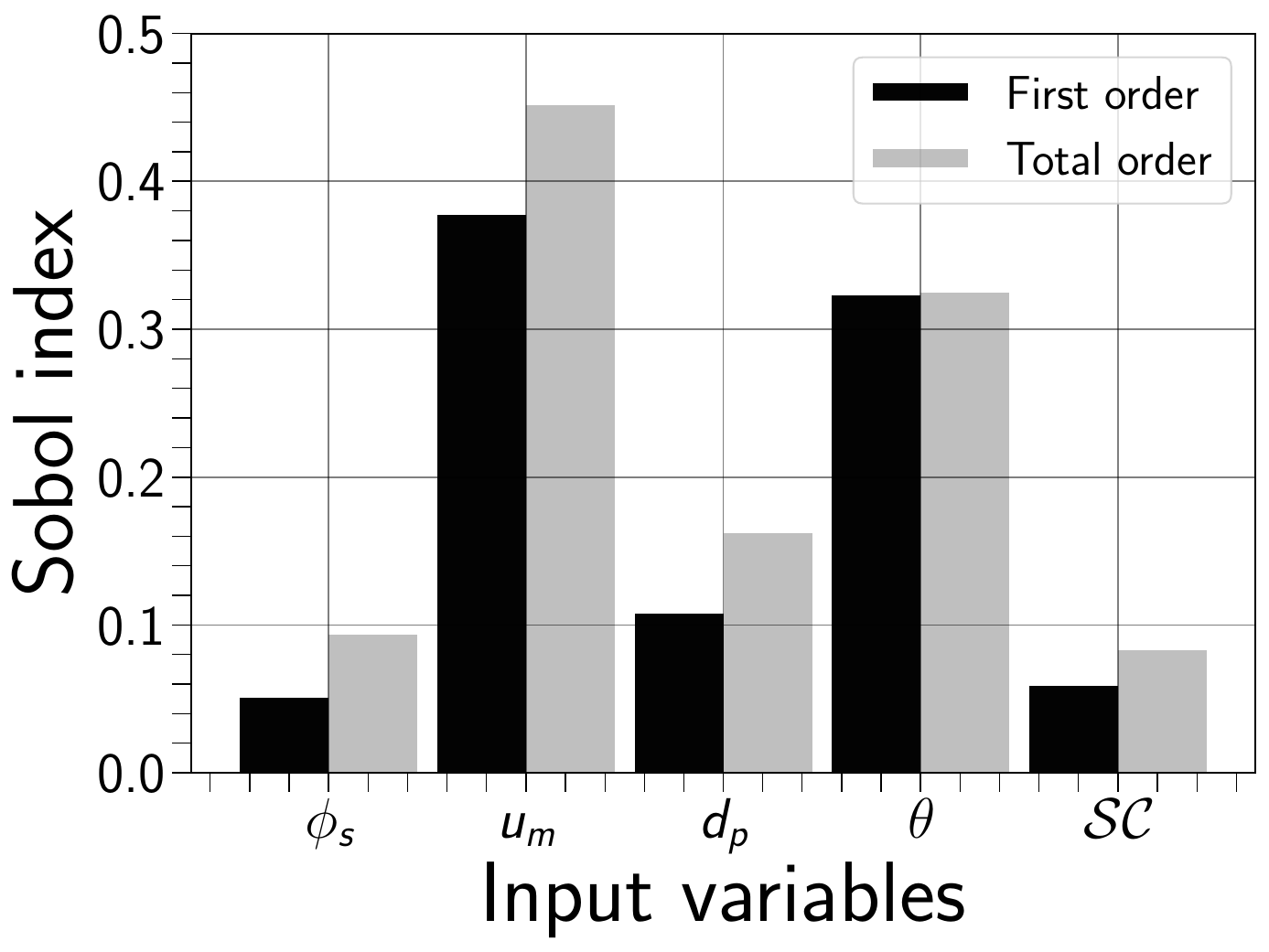}}\hfill
\subfloat[]{\includegraphics[width=0.45\textwidth]{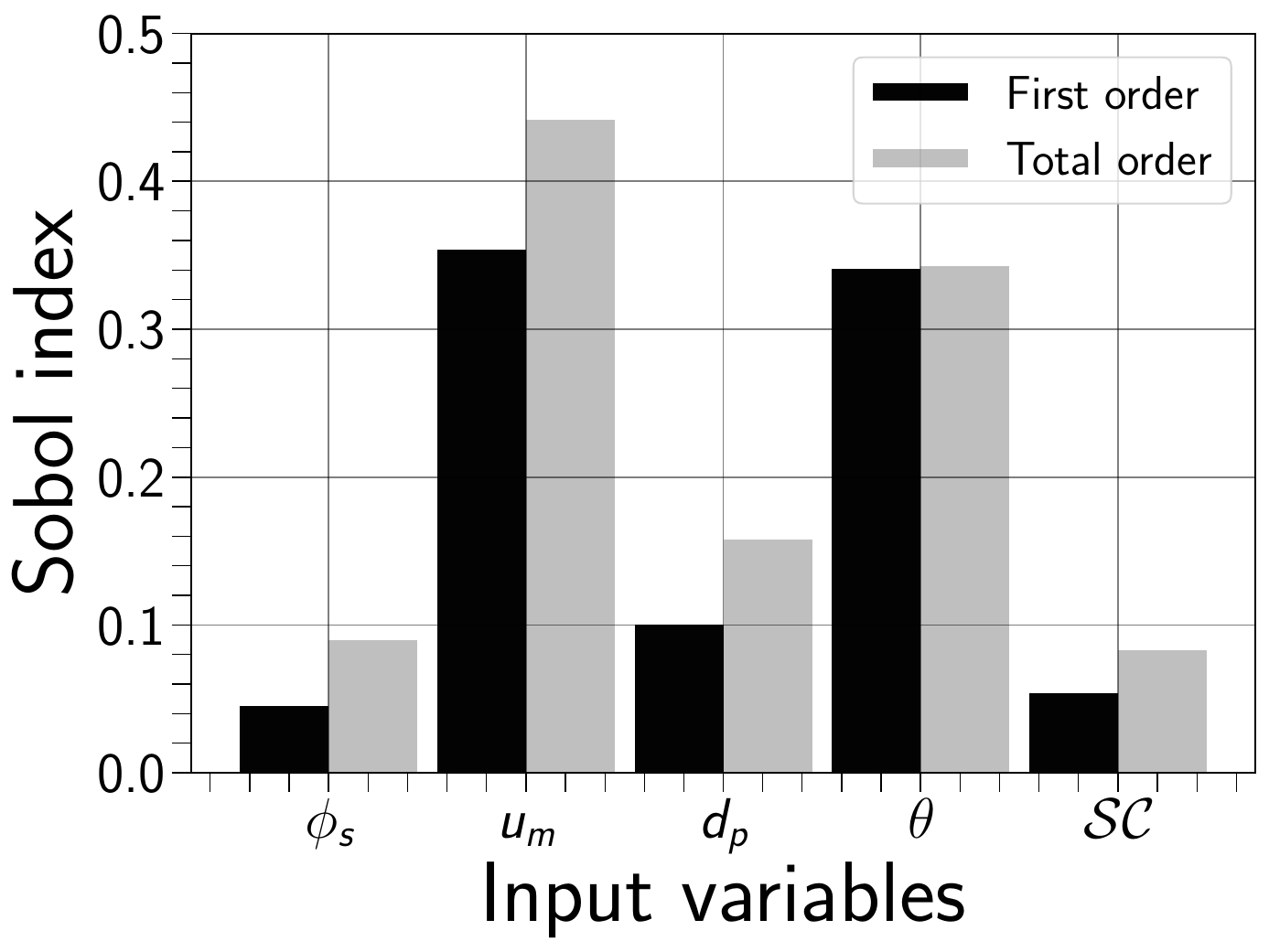}}\\
\caption{Barplots of first and total Sobol’ indices for the flow's pressure drop calculated based on: (a) One-dimensional data, (b) Two-dimensional data }
\label{fig:2D_P_agregé}
\end{figure}

The ability of the surrogate model to estimate the entire Probability Density Function (PDF) of the flow's pressure drop is examined.
This comparison is depicted in Fig.~\ref{fig:KS_POD_DP}, where the estimation of the complete PDF using Monte Carlo simulation is compared to that obtained using the PCE-POD surrogate model.
Sobol' indices are utilized to evaluate the sensitivity of slurry flow pressure.
The first and total aggregated Sobol' indices are shown in Fig.~\ref{fig:2D_P_agregé}.
The reduced model exhibits the same ranking of the most influential parameters as the full-order model.
The results clearly indicate that the velocity and pipe inclination have the greatest impact on the pressure drop, followed by particle size, particle concentration, and the specularity coefficient.
When considering the range of pipe inclination variations, it is concluded that it only affects the magnitude of the pressure drop and does not result in a pressure gain when the pipe is inclined downward.
The particle size influences the friction along the lower wall, and as the slurry flows through the pipeline, the different layers of particles interact with the component walls in varying ways.
Moreover, the layer with the highest particle concentration experiences greater frictional resistance against the wall, leading to a larger pressure drop compared to the layer with lower particle concentration.
Consequently, a pressure gradient forms across the slurry, where the pressure decreases as the particle concentration increases.
%
\section{Conclusion}
\label{general conclusion}
In this study, we have presented and discussed an efficient analysis of UQ for slurry flows problems.
Slurry flows are complex and highly nonlinear, making UQ studies valuable for gaining physical insights into the underlying processes.
Our work focuses on two-dimensional QoIs problems, where we have applied a reduced-order model to the two-dimensional Two-Fluid model Navier-Stokes equations.
The adopted strategy combines PCE with the POD technique, which allows to selectively retain the most relevant PCEs that best capture the underlying flow physics. Consequently, computational time is significantly reduced, enabling computations to be executed within a few minutes.
A summary of the proposed methodology developed in this work is presented in Fig.~\ref{fig:UQ-flowchart}.
\begin{figure}[ht!]
\centering
\includegraphics[width=\textwidth]{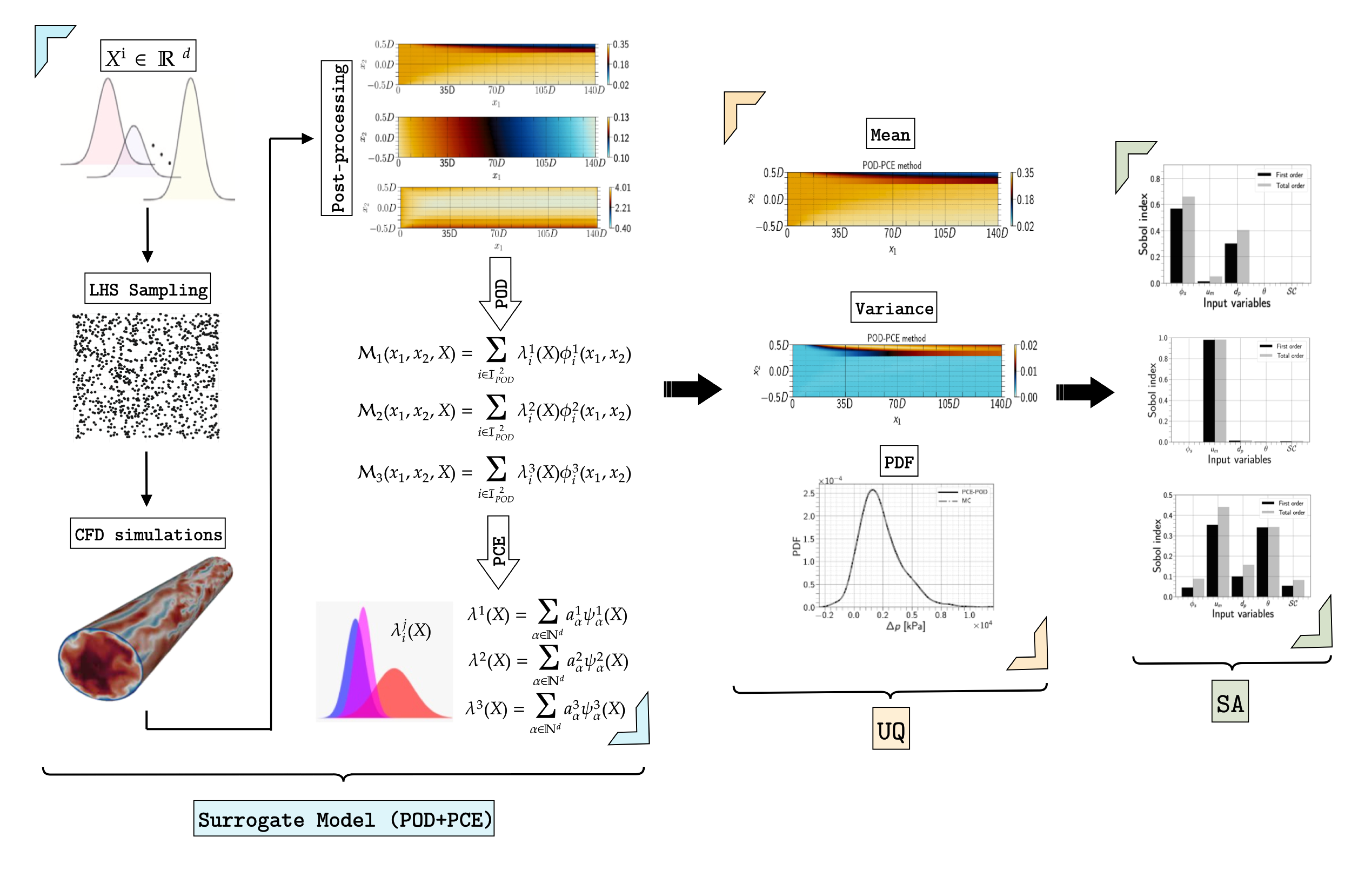}
\caption{The workflow of the UQ analyses based on POD-PCE technique.}
\label{fig:UQ-flowchart}
\end{figure}
The numerical results demonstrate that our developed reduced-order model is capable of producing acceptable results for statistical quantities of interest.
We compare the mean and variance distributions obtained from the reduced-order model with those computed using deterministic simulations, revealing that the variability of each QoI is well represented.
This comparison provides valuable information about the distribution of the outputs.

Our analysis also evaluates the influence of uncertain parameters on the QoIs, shedding light on the energy efficiency of slurry transport.
Notably, we observed a noticeable impact on the results when increasing the dimensionality of the QoIs from one to two dimensions.
Incorporating two-dimensional QoIs in our UQ and GSA study allows for a more robust and comprehensive assessment of uncertainties and sensitivities in complex slurry flow problems.
Specifically, the additional information gained regarding the sensitivity of our QoIs to input dimensions is crucial.
Parameters with high Sobol' indices play a significant role in ensuring accurate measurements and should be prioritized.
By determining appropriate measurement uncertainty for these parameters and understanding the level of relaxation in measurement sensitivity for others, engineers can mitigate nonlinear effects caused by stochastic measurement errors and ultimately reduce uncertainty.
Furthermore, the inherent uncertainty in field measurements can introduce uncertainty into CFD simulations aimed at predicting the specific energy consumption (SEC).
Prioritizing simulation parameters with high uncertainty levels helps reduce measurement uncertainty and enhances simulation reliability.
Therefore, controlling parameters that have a significant impact on the overall uncertainty of the simulation is important for improved control and optimization of slurry flow, leading to more reliable and accurate simulations.

Our statistical approach surpasses the classical "one-at-a-time" sensitivity analyses by employing GSA and UQ.
This approach considers the joint effects of multiple parameters, providing a more comprehensive understanding of slurry flows in pipelines, a complex system.
By demonstrating the effectiveness of these techniques, our study serves as a prototype, paving the way for future research in the field of slurry flow analysis and optimization.

\subsection*{Conflict of Interest}
The authors have no conflicts to disclose.

\subsection*{Data availability}
The data that support the findings of this study are available from the corresponding author upon reasonable request.

\subsection*{Acknowledgement}
This work was supported by OCP Group (Morocco). The authors gratefully acknowledge the support and computing resources from the African Supercomputing Center (ASCC) at UM6P (Morocco).
\appendix\normalsize
\section{Solver validation and verification}
\subsection{Numerical validation}
\begin{figure}[ht!]
\centering
\begin{subfigure}{0.49\textwidth}
\includegraphics[width=0.99\columnwidth]{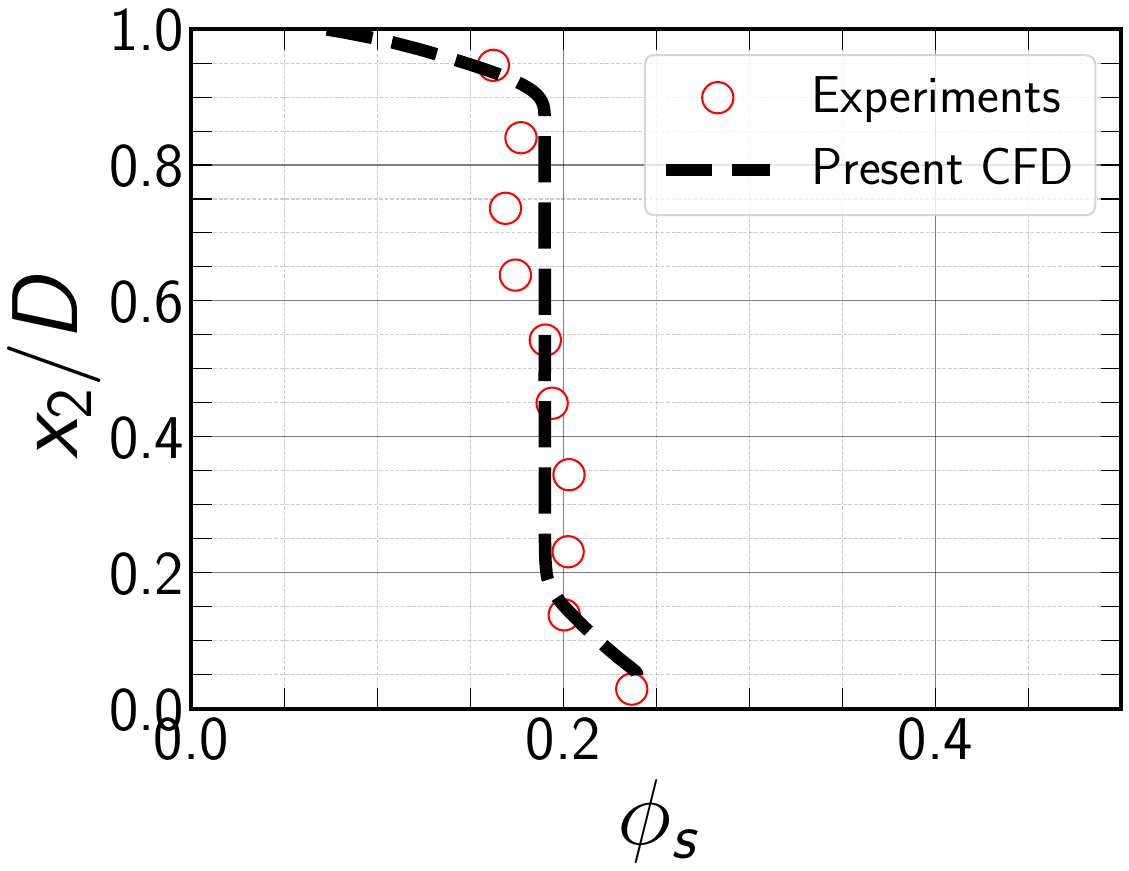}
\caption{}
\label{fig:phi=19}
\end{subfigure}
\begin{subfigure}{0.49\textwidth}
\includegraphics[width=0.99\columnwidth]{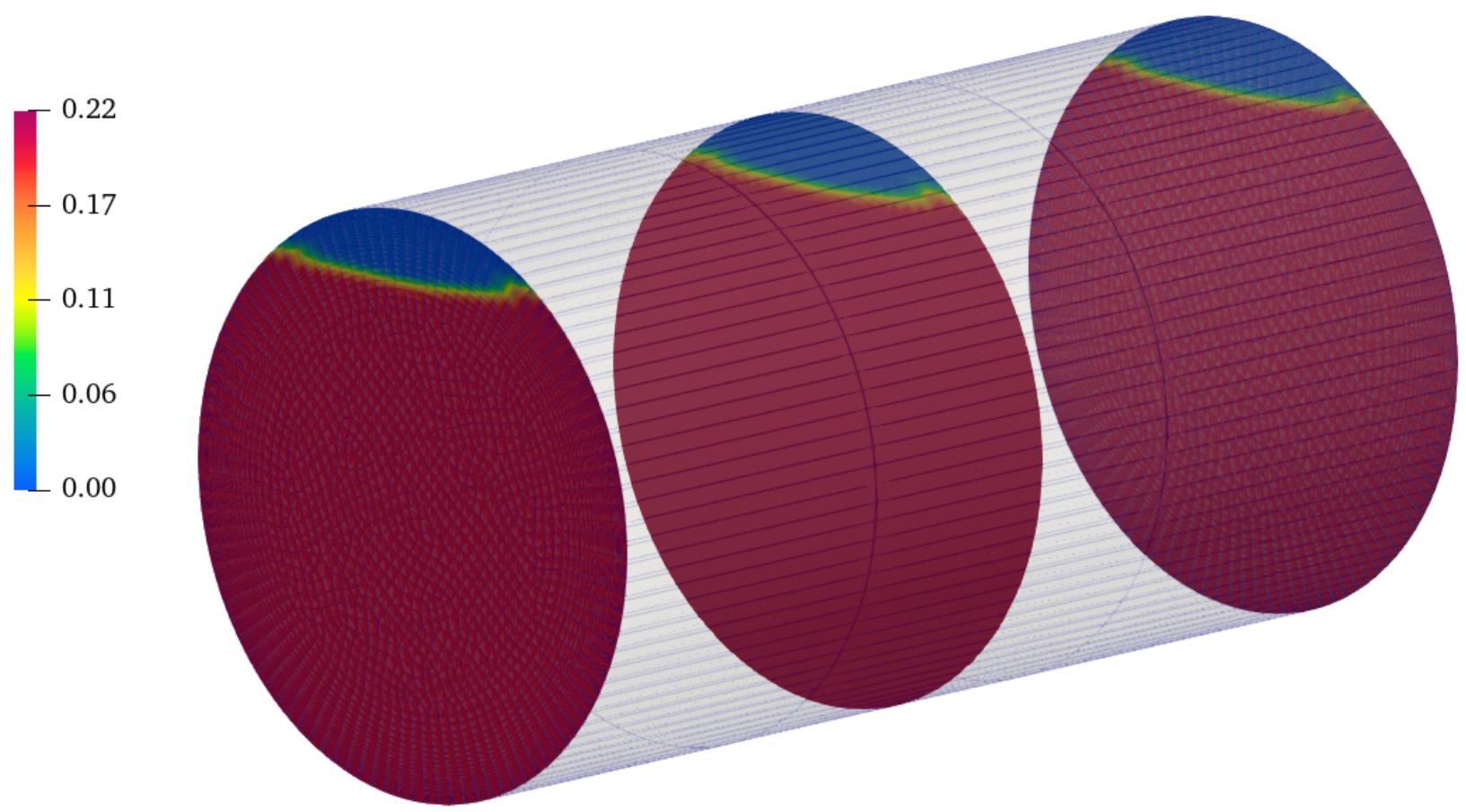}
\caption{}
\label{fig:Pipe_section_19}
\end{subfigure}
\caption{\color{red} Comparison of radial distributions of the volume fraction of sand particles $\varepsilon_{{s}}$: (a) $\mathrm{d}_{{p}}= 0.09$ mm and ${u_{m}}=3$ m/s, (b) Cross section at the pipe outlet.}
\label{fig:}
\end{figure}

\begin{figure}[ht!]
\centering
\begin{subfigure}{0.48\textwidth}
\includegraphics[width=0.99\textwidth]{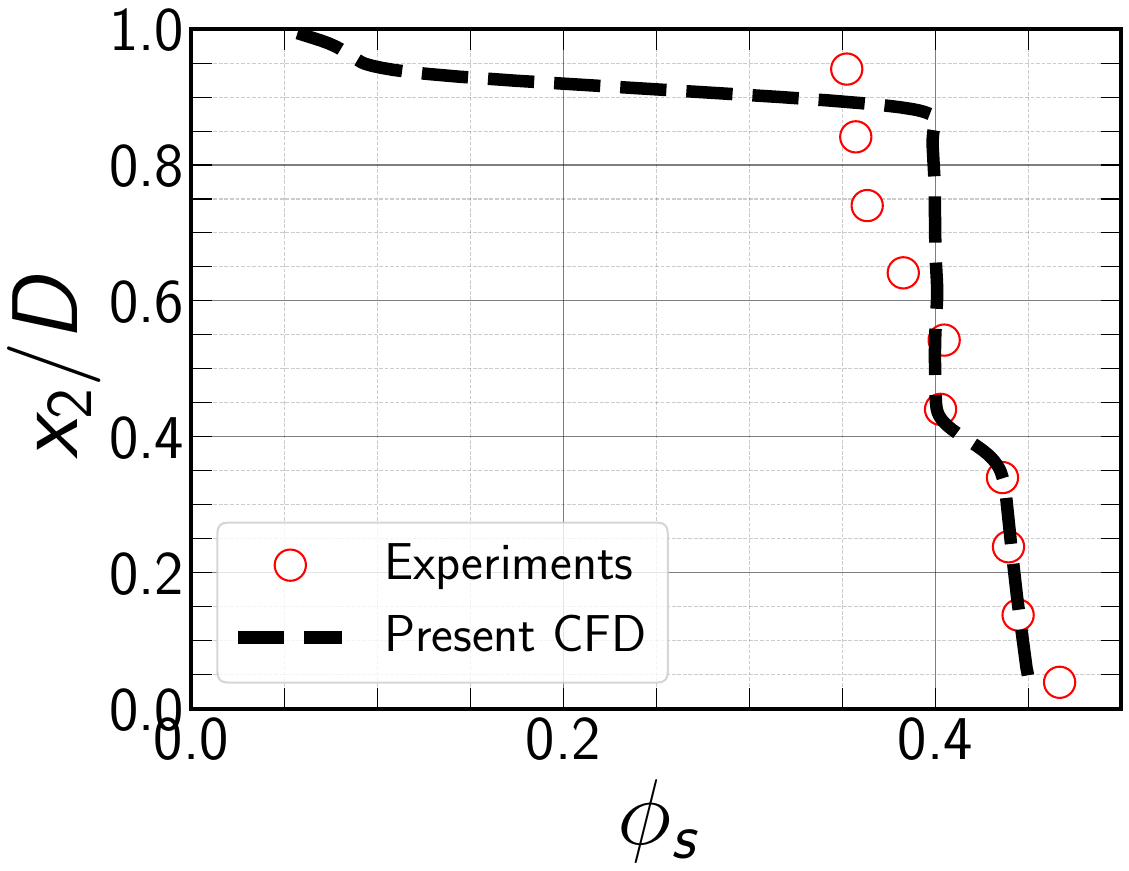}
\caption{}
\label{fig:phi=40}
\end{subfigure}
\begin{subfigure}{0.48\textwidth}
\includegraphics[width=0.99\textwidth]{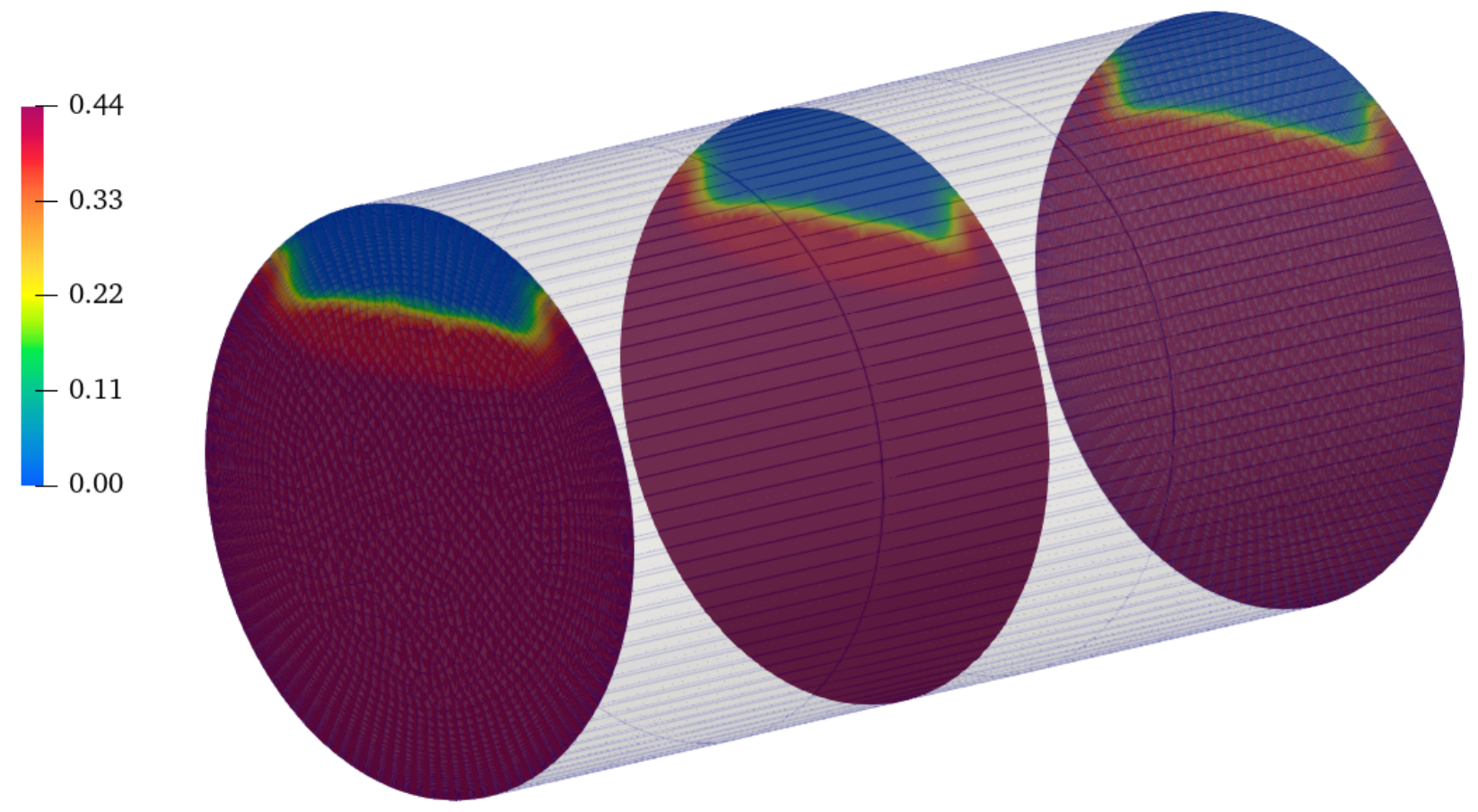}
\caption{}
\label{fig:Pipe_section_40}
\end{subfigure}
\caption{\color{red} Comparison of radial distributions of the volume fraction of sand particles $\phi_{{s}}$: (a) $\mathrm{d}_{{p}}= 0.27$ mm and ${u_{m}}=5.4$ m/s, (b) Cross-section at the pipe outlet}
\label{fig:}
\end{figure}

\begin{figure}[ht!]
\centering
\begin{subfigure}{0.49\textwidth}
\includegraphics[width=0.9\textwidth]{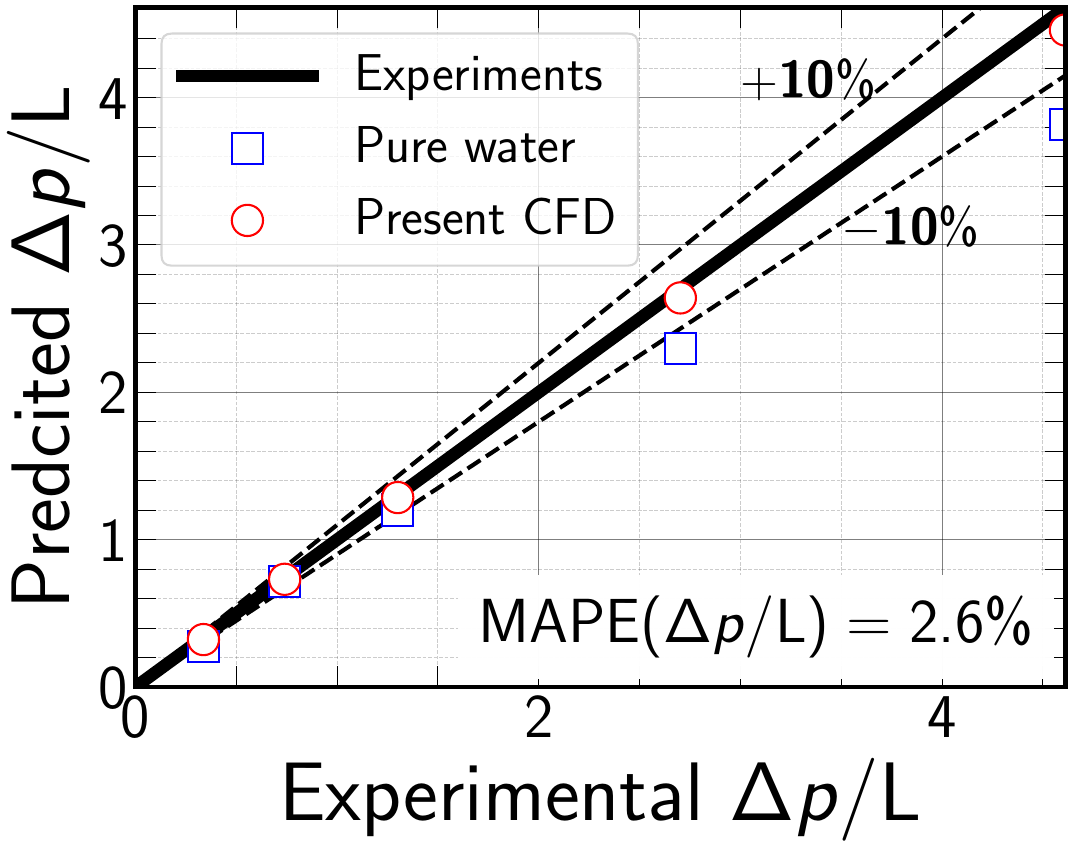}
\caption{$\phi_{{s}}=0.19$, $\mathrm{d}_{{p}}=0.09$ mm}
\label{fig:}
\end{subfigure}
\begin{subfigure}{0.49\textwidth}
\includegraphics[width=0.9\textwidth]{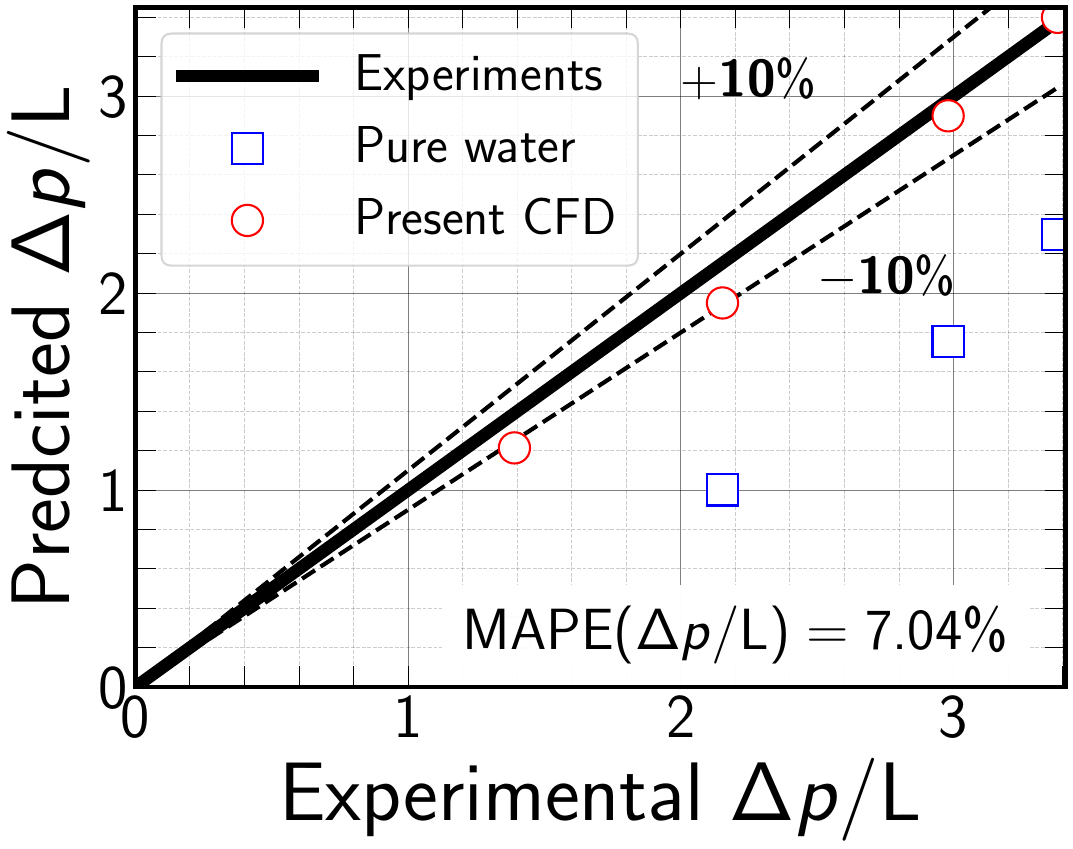}
\caption{$\phi_{{s}}=0.40$, $\mathrm{d}_{{p}}=0.27$ mm}
\label{fig:}
\end{subfigure}
\caption{\color{red} Predicted pressure drop per unit of length $\Delta {p}/L$ versus experimental data. Differences in relative errors between predictions and experiments are also displayed.}
\label{fig:pressureDist}
\end{figure}

%
The present modeling approach undergoes validation against experimental data obtained from a sand-water slurry outlined in the literature \citet{gillies2004modelling}. 
Comparisons between experimental and simulation data for particle concentrations at $\phi_{{s}} = 0.19$ and $\phi_{{s}} = 0.4$ along the vertical diameter direction across the horizontal pipe cross-section (cf. Fig.~\ref{fig:Pipe_section_19} and Fig.~\ref{fig:Pipe_section_40}) are depicted in Fig.~\ref{fig:phi=19} and Fig.~\ref{fig:phi=40}, respectively.
The numerical model's performance is further analyzed by comparing how the pressure drop per unit length changes with respect to the inlet velocity of the mixture, ${u}_{m}$, ranging from $2$ m/s to $8$ m/s.
The results are illustrated in Fig.~\ref{fig:pressureDist}.
Our CFD results show a significant alignment with the experimental data provided.
The overall mean relative errors in pressure drops remained below 10$\%$, a level widely deemed acceptable.
This proves the validity of the numerical model within the specified range of physical conditions.

\bibliography{main.bib}

\end{document}